\documentclass[12pt]{article}    
\usepackage{setspace}
\usepackage{hyperref}
\usepackage{graphics}
\usepackage{epsfig}
\usepackage{url}
\usepackage{amsmath}
\usepackage{mathrsfs}
\usepackage[utf8]{inputenc}
\usepackage{graphicx}
\usepackage{fancyhdr}
\usepackage[flushleft]{threeparttable}
\usepackage[nottoc]{tocbibind}
\usepackage[graphicx]{realboxes}
\usepackage{pdflscape}

\usepackage{amssymb}
\usepackage{setspace}
\usepackage{hyperref}
\usepackage{graphics}
\usepackage{epsfig}
\usepackage{url}
\usepackage{amsmath}
\usepackage{mathrsfs}
\usepackage{amssymb}

\usepackage{amsmath}
\usepackage{bm}
\usepackage{amsfonts}
\usepackage{amsmath}
\usepackage{mathrsfs}
\usepackage{multirow}
\numberwithin{equation}{section}
\usepackage{tablefootnote}
\usepackage{booktabs}
\usepackage{multirow}
\usepackage{siunitx}
\usepackage[flushleft]{threeparttable}
\usepackage{color}
\usepackage{hyperref}
\usepackage{graphicx}
\usepackage{graphics}
\usepackage{adjustbox}
\usepackage{titlesec}

\usepackage{amssymb}

\def \R {{\rm I\kern -2.2pt R\hskip 1pt}}

\newcommand{\qed}{\rule{0.5em}{1.5ex}}

\usepackage{color}
\begin{document}
	
	\begin{center} { \large \sc  A Note on Bayesian Inference for the Bivariate Pseudo-Exponential Data}\\
		\vskip 0.1in {\bf Banoth Veeranna}\footnote{veerusukya40@gmail.com}
		\\
		\vskip 0.1in 
		School of Mathematics and Statistics, University of Hyderabad, Hyderabad, India. \\

		

	\end{center}
	
	\begin{abstract} 
		 In this present work, we discuss the Bayesian inference for the bivariate pseudo-exponential distribution. Initially, we assume independent gamma priors and then pseudo-gamma priors for the pseudo-exponential parameters. We are primarily interested in deriving the posterior means for each of priors assumed and also comparing each of the posterior means with the maximum likelihood estimators. Finally, all the Bayesian analyses are illustrated with a simulation study and also illustrated with real-life applications.
	\end{abstract}
	
	\textbf{keywords:} Posterior distributions, Confidence Intervals, Pseudo-Exponential, Gamma Priors, Exponential distribution, marginal and conditional distributions.
	
	\bigskip

	\section{Introduction }  Bivariate models that are conditionally specified frequently offer valuable, adaptable models with a range of dependence patterns. In such cases, consideration should be given to what are recognized as pseudo models, see \cite{FF2006}-\cite{FF2013}. In the present note we are interested in the Bayesian analysis of the bivariate pseudo-exponential which was primarily discussed in \cite{BM2019}.  The bivariate pseudo-exponential models will have one exponential marginal and another conditional is also exponential. We start by going over the view on the bivariate pseudo-exponential model focusing on a few simplified sub-models. The Bayesian inference for such sub-models are explained in detail and each of the inferential aspects are illustrated with simulation study. Finally, we give a real life application of the presumed bivariate distribution.  A similar phenomenon with Poisson structure is discussed in \cite{am21} and on the Bayesian with pseudo priors in \cite{BM2019}.

\section{The Bivariate Pseudo-Exponential Distributions}
Let X and Y be a random variables with $X>0$ and $Y>0$ that could be used as a model for the lifespans of connected system components. The joint density is specified as a specific non-negative function on $(0, \infty) \times (0, \infty)$ that integrates to 1. In contrast, \cite{FF2007} suggest using one marginal distribution (let's say X) and the family of conditional densities of the other variable (Y) given X=x, for every x, see \cite{BEJ99}, \cite{FF2006}. Therefore, the joint density of (X, Y) will have the following form, if h(x) denotes the density of X for each x, and $g_x(y)$ indicates the conditional density of Y given that X=x, then
\begin{equation}\label{eq1}
f_{X,Y}(x,y) = h(x)g_x(y)I(x>0, y>0).
\end{equation}
In general, the pseudo-exponential distributions (inside the context of Filus and Filus) the equivalent to the case where h and $g_x$'s have exponential densities. As a result, the joint density of a bivariate pseudo-exponential distribution is one where
\begin{equation}\label{eq2}
X \sim Exp(\theta_1),
\end{equation}
and for each $x>0$,
\begin{equation}\label{eq3}
Y|X=x \sim Exp(\theta(x)),
\end{equation}
where $\theta_1>0$ and $\theta(x)>0$ $\forall$ $x$. Except for measurability, there are no constraints on the structure of the function $\theta(x)$. The corresponding joint density is given by \begin{equation}\label{eq4}
f_{X,Y}(x,y) = \theta_1e^{-\theta_1x}\theta(x)e^{-\theta(x)y}I(x>0, y>0)
\end{equation}
where the positive function $\theta(x)$ is quite arbitrary. Suppose,
that $\theta(x) = \theta_2 + \theta_3x$, then the corresponding joint p.d.f. is given by
\begin{equation}\label{eq5}
f_{X,Y}(x,y) = \theta_1e^{-\theta_1x}(\theta_2 + \theta_3x)e^{-(\theta_2 + \theta_3x)y}I(x>0, y>0).
\end{equation}
where $\theta_i>0$, i=1,2,3.

\subsection{Notes on priors for pseudo-exponential data} \label{subsec:statistical-summaries}
The joint density function for a bivariate pseudo-exponential distribution is of the form equation (\ref{eq5}).\\
Consequently, the likelihood function for a sample of size n from this distribution is given by
\begin{equation}\label{eq6}
\begin{split}
L(\theta_1,\theta_2,\theta_3|x,y) &= \theta^n_1e^{-\theta_1\sum^n_{i=1}{x_i}} (\prod^n_{i=1}(\theta_2 + \theta_3{x_i})) \\
& e^{-\theta_2\sum^n_{i=1}{y_i} - \theta_3\sum^n_{i=1}{x_iy_i}}
\end{split}
\end{equation}
Note that this likelihood factors as follows:
\begin{equation}\label{eq7}
\begin{split}
L(\theta_1,\theta_2,\theta_3|x,y) &= \Big\{\theta^n_1e^{-\theta_1\sum^n_{i=1}{x_i}}\Big\} \\
& \Big\{(\prod^n_{i=1}(\theta_2 + \theta_3{x_i})) e^{-\theta_2\sum^n_{i=1}{y_i} - \theta_3\sum^n_{i=1}{x_iy_i}}\Big\}
\end{split}
\end{equation}
In the first factor only involves the parameter $\theta_1$, while the second factor involves the parameters $\theta_2$ and $\theta_3$. We will call $\theta_1$ the marginal parameter, and $\theta_2$ and $\theta_3$ will be called conditional parameters. This factorization will prove to be important in Bayesian inference for this model, as discuss below. Because of the factorization we will know that if a priori $\Tilde{\theta_1}$ and $(\Tilde{\theta_2}, \Tilde{\theta_3})$ are independent, then they will be independent in posteriori also.\\
Now, to the pseudo-exponential distribution defined in equation (\ref{eq5}), it is simply assume a gamma prior for the marginal parameter $\Tilde{\theta_1}$. i.e., a priori we assume that 
\begin{equation}\label{eq8}
\Tilde{\theta_1} \sim \Gamma(\alpha_1,\beta_1), 
\end{equation}
where $0 < \alpha_1 < \infty$ and $0 < \beta_1 < \infty$.\\
Therefore, the priori density for $\Tilde{\theta_1}$ is of the form
\begin{equation*}
f_{\Tilde{\theta_1}}(\theta_1) = \frac{\beta^{\alpha_1}_1\theta_1^{\alpha_1-1}e^{-\beta_1\theta_1}}{\Gamma(\alpha_1)};~~~~~~~~ 0<\theta_1<\infty.
\end{equation*}
The choice of  a priori joint density of $(\Tilde{\theta_2}, \Tilde{\theta_3})$ that will be independent of $\Tilde{\theta_1}$ is not so obvious. It is clear whether it is possible to choose such a prior that will be "Conjugate" with the second factor in equation (\ref{eq7}). The joint posterior of $(\Tilde{\theta_2}, \Tilde{\theta_3})$ will need to be dealt with numerically.

\section{Independent priors} \label{sec:Independents}
Consider an a priori joint density in which all three parameters are independent, and each has a gamma distribution. Thus we have
\begin{equation}\label{eq9}
f_{\Tilde{\theta_1},\Tilde{\theta_2},\Tilde{\theta_3}}(\theta_1,\theta_2,\theta_3) = \prod^3_{i=1}\frac{\beta_i^{\alpha_i}\theta_i^{\alpha_i-1}e^{-\beta_i\theta_i}}{\Gamma(\alpha_i)}
\end{equation}
where $0 < \theta_i, \alpha_i, \beta_i <\infty$, for i=1,2,3.\\
The kernel of the posterior which is the product of the kernel of the factored likelihood (\ref{eq7}) and the kernel of the prior (\ref{eq9}) is of the form 
\begin{equation}\label{eq10}
\begin{split}
ker(f_{\Tilde{\theta_1},\Tilde{\theta_2},\Tilde{\theta_3}|\underline{X},\underline{Y}}(\theta_1,\theta_2,\theta_3|\underline{x},\underline{y})) \\
= \Big \{\theta_1^{\alpha_1+n-1}e^{-\theta_1(\beta_1+\sum^n_{i=1}x_i)} \Big \}\\
* \Big \{(\prod^n_{i=1}(\theta_2+\theta_3{x_i}))\theta_2^{\alpha_2-1}\theta_3^{\alpha_3-1}e^{-\theta_2(\beta_2+\sum^n_{i=1}{y_i})} \\
* e^{-\theta_3(\beta_3+\sum^n_{i=1}{x_iy_i})}\Big \}
\end{split}
\end{equation}
From the first factor in equation (\ref{eq10}) we recognise that a posteriori $\Tilde{\theta_1}$ has a gamma distribution, i.e.,
\begin{equation}\label{eq11}
\Tilde{\theta_1}|\underline{X}=\underline{x},\underline{Y}=\underline{y} \sim \Gamma(\alpha_1+n, \beta_1+\sum^n_{i=1}{x_i})
\end{equation}
The second factor is the kernel of the posterior density of $(\Tilde{\theta_2}, \Tilde{\theta_3})$. It will need to be dealt with numerically.
\subsection{Sub-Model-I($\theta_2=\theta_3$):}
We first focus on the the sub-model of equation (5) obtained by equating $\theta_2$ and $\theta_3$. The model is given by
\begin{equation}\label{eq12}
f_{X,Y}(x,y) = \theta_1e^{-\theta_1x}\theta_3(1+x)e^{-\theta_3(1+x)y}I(x>0,y>0).
\end{equation}
The likelihood function for a sample of size n from this distribution is 
\begin{equation*}
L(\theta_1, \theta_3) = \theta_1^n e^{-\theta_1\sum^n_{i=1}x_i}\theta_3^n(\prod^n_{i=1}(1+x_i)) e^{-\theta_3(\sum^n_{i=1}y_i+\sum^n_{i=1}x_iy_i)}
\end{equation*}
The likelihood factors as follows:
\begin{equation}\label{eq13}
\begin{split}
L(\theta_1, \theta_3) & = \Big \{ \theta_1^n e^{-\theta_1\sum^n_{i=1}x_i} \Big \} \\
& \Big \{\theta_3^n(\prod^n_{i=1}(1+x_i)) e^{-\theta_3(\sum^n_{i=1}y_i+\sum^n_{i=1}x_iy_i)} \Big \}
\end{split}
\end{equation}
For such a conjugate joint prior density for two parameters in the model, we can take one with independent gamma marginals. Thus
\begin{equation}\label{eq14}
f_{\Tilde{\theta_1},\Tilde{\theta_3}}(\theta_1,\theta_3) = \frac{\beta^{\alpha_1}_1\theta_1^{\alpha_1-1}e^{-\beta_1\theta_1}}{\Gamma(\alpha_1)}. \frac{\beta^{\alpha_3}_3\theta_3^{\alpha_3-1}e^{-\beta_3\theta_3}}{\Gamma(\alpha_3)}
\end{equation}
The kernel of the posterior density, which is the product of the kernel of the likelihood factors (\ref{eq13}) and the kernel of the joint prior (\ref{eq14}) is of the form
\begin{equation}\label{eq15}
\begin{split}
Ker(f_{\Tilde{\theta_1},\Tilde{\theta_3}|\underline{X},\underline{Y}}(\theta_1,\theta_3|\underline{x},\underline{y})) & = \Big \{ \theta_1^{\alpha_1+n-1}e^{-\theta_1(\beta_1+\sum^n_{i=1}x_i)} \Big \} \\
& * \Big \{ \theta_3^{\alpha_3+n-1}e^{-\theta_3(\beta_3+\sum^n_{i=1}y_i+\sum^n_{i=1}x_iy_i)} \Big \}
\end{split}
\end{equation}
A posteriori the two parameters have independent gamma distributions. Thus
\begin{equation}\label{eq16}
\Tilde{\theta_1}|\underline{X}=\underline{x},\underline{Y}=\underline{y} \sim \Gamma(\alpha_1+n, \beta_1+\sum^n_{i=1}{x_i}),
\end{equation}
and, independently, 
\begin{equation}\label{eq17}
\Tilde{\theta_3}|\underline{X}=\underline{x},\underline{Y}=\underline{y} \sim \Gamma(\alpha_3+n, \beta_3+\sum^n_{i=1}{y_i}+\sum^n_{i=1}{x_iy_i}).
\end{equation}
The squared error loss estimates of the parameters (the posterior means) are thus,
\begin{equation*}
\hat{\theta_1}^{(B)} = \frac{\alpha_1+n}{\beta_1+\sum^n_{i=1}{x_i}}
\end{equation*}
and 
\begin{equation*}
\hat{\theta_3}^{(B)} = \frac{\alpha_3+n}{\beta_3+\sum^n_{i=1}{y_i}+\sum^n_{i=1}{x_iy_i}}
\end{equation*}
If we choose to an improper prior with $\alpha_1=\alpha_3=\beta_1=\beta_3=0$, then the resulting Bayes estimates coincide with the corresponding maximum likelihood estimates(M.L.E's).

\begin{figure}[htp]\label{fig1}
	\centering
	\includegraphics[width=7.5cm]{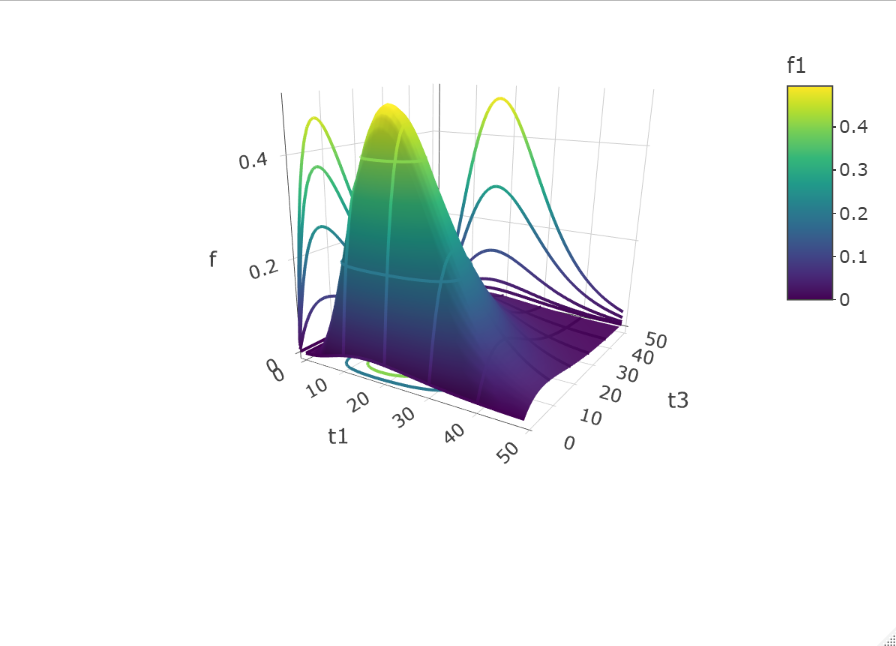}
	\caption{Density plot of independent gamma prior with hyper-parameter values are $\alpha_1=2, \alpha_3=4, \beta_1=2, \beta_3=3$ and parameter values $\theta_1=2, \theta_3=5$ with sample of size 30 of the sub-model-I.}
	\label{fig:Density Plots}
\end{figure}

\begin{figure}[htp]\label{fig2}
	\centering
	\includegraphics[width=7.5cm]{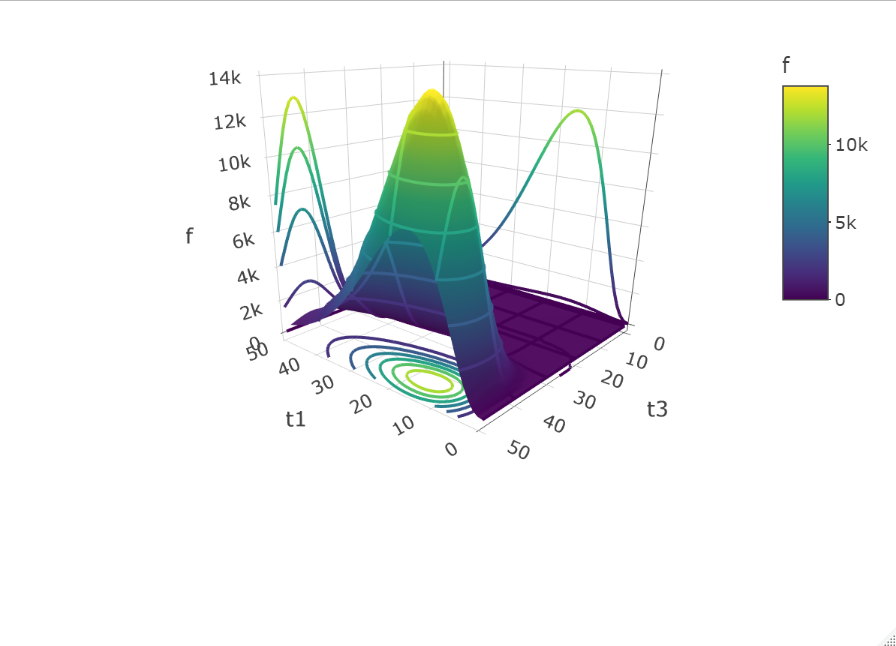}
	\caption{Density plot of posterior (independent gamma prior) with hyper-parameter values are $\alpha_1=2, \alpha_3=4, \beta_1=2, \beta_3=3$ and parameter values $\theta_1=2, \theta_3=5$ with sample of size 30 of the sub-model-I.}
	\label{fig:Density Plots}
\end{figure}
\newpage
\subsection*{A Predictive Model for sub-model I} Remember that our data \textbf{X} follow the distribution $f(\textbf{X}|\theta)$ for a fixed value of $\theta$. However, since the actual value of $\theta$ is unknown, we should take an average of all potential values in order to understand the distribution of $\textbf{X}$.
\par Now, $\underline{\boldsymbol{X}} \sim Exponential(\theta_1)$, and $\theta_1 \sim Gamma(\alpha_1, ~\beta_1)$ then the Posterior distribution of $\Tilde{\theta_1}|\underline{X}=\underline{x},\underline{Y}=\underline{y} \sim \Gamma(\alpha_1+n,  \beta_1+\sum^n_{i=1}x_i)$. \\
Therefore the Posterior Predictive distribution is:\\
\begin{equation*}
f(X_N|\underline{X},\underline{Y}) = \int_0^{\infty}f(X_N|\theta_1)f(\Tilde{\theta_1}|\underline{X},\underline{Y})d\theta_1
\end{equation*}
(Since $X_N$ is independent of the sample data $\boldsymbol{X}$.)
\begin{equation*}
\begin{split}
&= \int_0^{\infty}\left(\theta_1 e^{-\theta_1X_N}\right)\left(\frac{(\beta_1+\sum^n_{i=1}x_i)^{\alpha_1+n}}{\Gamma(\alpha_1+n)
}\theta_1^{\alpha_1+n-1}e^{-(\beta_1+\sum^n_{i=1}x_i)\theta_1}\right)d\theta_1 \\
&= \left(\frac{(\beta_1+\sum^n_{i=1}x_i)^{\alpha_1+n}}{\Gamma(\alpha_1+n)}\right)\int_0^{\infty}\theta_1^{\alpha_1+n+1}e^{-(\beta_1+\sum^n_{i=1}x_i+X_N)\theta_1}d\theta_1\\
&= \left(\frac{(\beta_1+\sum^n_{i=1}x_i)^{\alpha_1+n}}{\Gamma(\alpha_1+n)}\right)\left(\frac{\Gamma(\alpha_1+n+1)}{(\beta_1+\sum^n_{i=1}x_i+X_N)^{\alpha_1+n+1}}\right)\\
& f(X_N|\underline{X},\underline{Y}) = \frac{(\alpha_1+n)(\beta_1+\sum^n_{i=1}x_i)^{(\alpha_1+n)}}{(\beta_1+\sum^n_{i=1}x_i+X_N)^{(\alpha_1+n+1)}}.
\end{split}
\end{equation*}
This is the posterior predictive distribution for a new data point $X_N$. Now, which is follows a Pareto distribution of the second kind, with mean $\frac{\beta_1+\sum^n_{i=1}x_i}{\alpha_1+n-1}$ and variance is $\frac{(\beta_1+\sum^n_{i=1}x_i)^2(\alpha_1+n)}{(\alpha_1+n-1)^2(\alpha_1+n-2)}$. \\
Similarly, $\underline{Y}|\underline{X} \sim Exponential(\theta_3(1+x))$, and $\theta_3 \sim Gamma(\alpha_3, \beta_3)$. Then the Posterior distribution $\Tilde{\theta_3}|\underline{X}=\underline{x},\underline{Y}=\underline{y} \sim Gamma(\alpha_3+n,  \beta_3+\sum^n_{i=1}y_i+\sum^n_{i=1}x_iy_i)$.
Therefore the Posterior Predictive distribution is:\\
\begin{equation*}
f(X_N|\underline{X},\underline{Y}) = \int_0^{\infty}f(X_N|\theta_3)f(\Tilde{\theta_3}|\underline{X},\underline{Y})d\theta_3
\end{equation*}\\
\begin{equation*}
\begin{split}
&= \int_0^{\infty}\left(\theta_3 e^{-\theta_3X_N}\right)\{\frac{(\beta_3+\sum^n_{i=1}{y_i}+\sum^n_{i=1}{x_iy_i})^{\alpha_3+n}}{\Gamma(\alpha_3+n)} \\
& ~~~~~~~~~ \theta_3^{\alpha_3+n-1}e^{-(\beta_3+\sum^n_{i=1}{y_i}+\sum^n_{i=1}{x_iy_i})\theta_3}\}d\theta_3 \\
\end{split}
\end{equation*}
therefore
\begin{equation*}
f(X_N|\underline{X},\underline{Y}) = \frac{(\alpha_3+n)(\beta_3+\sum^n_{i=1}{y_i}+\sum^n_{i=1}{x_iy_i})^{(\alpha_3+n)}}{(\beta_3+\sum^n_{i=1}y_i+\sum^n_{i=1}x_iy_i+X_N)^{(\alpha_3+n+1)}}.
\end{equation*}
It is follows a Pareto distribution of the second kind, with mean $\frac{\beta_3+\sum^n_{i=1}{y_i}+\sum^n_{i=1}{x_iy_i}}{\alpha_3+n}$ and variance is $\frac{(\beta_3+\sum^n_{i=1}{y_i}+\sum^n_{i=1}{x_iy_i})^2(\alpha_3+n)}{(\alpha_3+n-1)^2(\alpha_3+n-2)}$.

\subsection{Sub-Model-II($\theta_2=0$):}
Consider the sub-model-II obtained by setting $\theta_2=0$. The model is thus of the form
\begin{equation}\label{eq18}
f_{X,Y}(x,y) = \theta_1e^{-\theta_1x}\theta_3xe^{-\theta_3xy}I(x>0,y>0).
\end{equation}
The likelihood function for a sample of size n from this distribution is 
\begin{equation*}
L(\theta_1, \theta_3) = \theta_1^n e^{-\theta_1\sum^n_{i=1}x_i}\theta_3^n(\prod^n_{i=1}x_i) e^{-\theta_3\sum^n_{i=1}x_iy_i}
\end{equation*}
The likelihood factors as follows:
\begin{equation}\label{eq19}
L(\theta_1, \theta_3) = \Big \{ \theta_1^n e^{-\theta_1\sum^n_{i=1}x_i} \Big \} \Big \{\theta_3^n(\prod^n_{i=1}x_i) e^{-\theta_3\sum^n_{i=1}x_iy_i} \Big \}
\end{equation}
If we use a prior with independent gamma marginals such as
\begin{equation}\label{eq20}
f_{\Tilde{\theta_1},\Tilde{\theta_3}}(\theta_1,\theta_3) = \frac{\beta^{\alpha_1}_1\theta_1^{\alpha_1-1}e^{-\beta_1\theta_1}}{\Gamma(\alpha_1)}. \frac{\beta^{\alpha_3}_3\theta_3^{\alpha_3-1}e^{-\beta_3\theta_3}}{\Gamma(\alpha_3)}
\end{equation}
The kernel of the posterior density, which is the product of the kernel of the likelihood factors (\ref{eq19}) and the kernel of the joint prior (\ref{eq20}) is given by
\begin{equation}\label{eq21}
\begin{split}
Ker(f_{\Tilde{\theta_1},\Tilde{\theta_3}|\underline{X},\underline{Y}}(\theta_1,\theta_3|\underline{x},\underline{y})) & = \Big \{ \theta_1^{\alpha_1+n-1}e^{-\theta_1(\beta_1+\sum^n_{i=1}x_i)} \Big \} \\
& * \Big \{ \theta_3^{\alpha_3+n-1}e^{-\theta_3(\beta_3+\sum^n_{i=1}x_iy_i)} \Big \}
\end{split}
\end{equation}
A posteriori the two parameters have independent gamma distributions. Thus
\begin{equation}\label{eq22}
\Tilde{\theta_1}|\underline{X}=\underline{x},\underline{Y}=\underline{y} \sim \Gamma(\alpha_1+n, \beta_1+\sum^n_{i=1}{x_i}),
\end{equation}
and, independently, 
\begin{equation}\label{eq23}
\Tilde{\theta_3}|\underline{X}=\underline{x},\underline{Y}=\underline{y} \sim \Gamma(\alpha_3+n, \beta_3+\sum^n_{i=1}{x_iy_i}).
\end{equation}
The squared error loss estimates of the parameters (the posterior means) are thus,
\begin{equation*}
\hat{\theta_1}^{(B)} = \frac{\alpha_1+n}{\beta_1+\sum^n_{i=1}{x_i}}
\end{equation*}
and 
\begin{equation*}
\hat{\theta_3}^{(B)} = \frac{\alpha_3+n}{\beta_3+\sum^n_{i=1}{x_iy_i}}
\end{equation*}
If we choose to an improper prior with $\alpha_1=\alpha_3=\beta_1=\beta_3=0$, then the resulting Bayes estimates coincide with the corresponding maximum likelihood estimates(M.L.E's).
\begin{figure}[htp]\label{fig3}
	\centering
	\includegraphics[width=7.5cm]{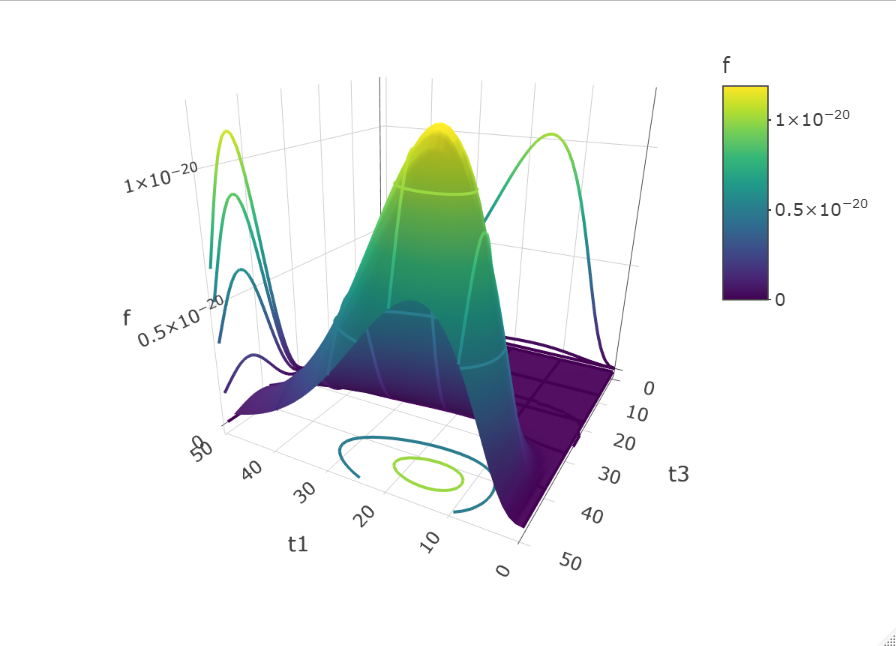}
	\caption{Density plot of posterior (independent gamma prior) with hyper-parameter values are $\alpha_1=2, \alpha_3=4, \beta_1=2, \beta_3=3$ and parameter values $\theta_1=2, \theta_3=5$ with sample of size 30 of the sub-model-II.}
	\label{fig:Density Plots}
\end{figure}
\subsubsection{A Predictive Model for sub-model II:}
Remember that our data \textbf{X} follow the distribution $f(\textbf{X}|\theta)$ for a fixed value of $\theta$. However, since the actual value of $\theta$ is unknown, we should take an average of all potential values in order to understand the distribution of $\textbf{X}$.\\
Now, $\underline{\boldsymbol{X}} \sim Exponential(\theta_1)$, and $\theta_1 \sim Gamma(\alpha_1, ~\beta_1)$ then the Posterior distribution of $\Tilde{\theta_1}|\underline{X}=\underline{x},\underline{Y}=\underline{y} \sim \Gamma(\alpha_1+n,  \beta_1+\sum^n_{i=1}x_i)$. \\
Therefore the Posterior Predictive distribution is:\\
\begin{equation*}
f(X_N|\underline{X},\underline{Y}) = \int_0^{\infty}f(X_N|\theta_1)f(\Tilde{\theta_1}|\underline{X},\underline{Y})d\theta_1
\end{equation*}
(Since $X_N$ is independent of the sample data $\boldsymbol{X}$.)
\begin{align*}
\begin{split}
&= \int_0^{\infty}\left(\theta_1 e^{-\theta_1X_N}\right)\left(\frac{(\beta_1+\sum^n_{i=1}x_i)^{\alpha_1+n}}{\Gamma(\alpha_1+n)
}\theta_1^{\alpha_1+n-1}e^{-(\beta_1+\sum^n_{i=1}x_i)\theta_1}\right)d\theta_1 \\
&= \left(\frac{(\beta_1+\sum^n_{i=1}x_i)^{\alpha_1+n}}{\Gamma(\alpha_1+n)}\right)\int_0^{\infty}\theta_1^{\alpha_1+n+1}e^{-(\beta_1+\sum^n_{i=1}x_i+X_N)\theta_1}d\theta_1\\
&= \left(\frac{(\beta_1+\sum^n_{i=1}x_i)^{\alpha_1+n}}{\Gamma(\alpha_1+n)}\right)\left(\frac{\Gamma(\alpha_1+n+1)}{(\beta_1+\sum^n_{i=1}x_i+X_N)^{\alpha_1+n+1}}\right)\\
& f(X_N|\underline{X},\underline{Y}) = \frac{(\alpha_1+n)(\beta_1+\sum^n_{i=1}x_i)^{(\alpha_1+n)}}{(\beta_1+\sum^n_{i=1}x_i+X_N)^{(\alpha_1+n+1)}}.
\end{split}
\end{align*}
This is the posterior predictive distribution for a new data point $X_N$. Now, which is follows a Pareto distribution of the second kind, with mean $\frac{\beta_1+\sum^n_{i=1}x_i}{\alpha_1+n-1}$ and variance is $\frac{(\beta_1+\sum^n_{i=1}x_i)^2(\alpha_1+n)}{(\alpha_1+n-1)^2(\alpha_1+n-2)}$. \\
Similarly, $\underline{Y}|\underline{X} \sim Exponential(\theta_3x)$, and $\theta_3 \sim Gamma(\alpha_3, \beta_3)$. Then the Posterior distribution $\Tilde{\theta_3}|\underline{X}=\underline{x},\underline{Y}=\underline{y} \sim Gamma(\alpha_3+n,  \beta_3+\sum^n_{i=1}x_iy_i)$.
Therefore the Posterior Predictive distribution is:\\
\begin{equation*}
f(X_N|\underline{X},\underline{Y}) = \int_0^{\infty}f(X_N|\theta_3)f(\Tilde{\theta_3}|\underline{X},\underline{Y})d\theta_3
\end{equation*}
\begin{equation*}
= \int_0^{\infty}\left(\theta_3 e^{-\theta_3X_N}\right)\left(\frac{(\beta_3+\sum^n_{i=1}{x_iy_i})^{\alpha_3+n}}{\Gamma(\alpha_3+n)
}\theta_3^{\alpha_3+n-1}e^{-(\beta_3+\sum^n_{i=1}{x_iy_i})\theta_3}\right)d\theta_3
\end{equation*}
therefore
\begin{equation*}
f(X_N|\underline{X},\underline{Y}) = \frac{(\alpha_3+n)(\beta_3+\sum^n_{i=1}{x_iy_i})^{(\alpha_3+n)}}{(\beta_3+\sum^n_{i=1}y_i+\sum^n_{i=1}x_iy_i+X_N)^{(\alpha_3+n+1)}}.
\end{equation*}
It is follows a Pareto distribution of the second kind, with mean $\frac{\beta_3+\sum^n_{i=1}{x_iy_i}}{\alpha_3+n}$ and variance is $\frac{(\beta_3+\sum^n_{i=1}{x_iy_i})^2(\alpha_3+n)}{(\alpha_3+n-1)^2(\alpha_3+n-2)}$.

\section{Pseudo-Gamma priors}
In the following sections we will be discussing bivariate pseudo-gamma priors and their applications in more details for each of the sub-model I \& II.

\subsection{Sub-Model I ($\theta_2=\theta_3$):}
Consider the sub-model, specified in equation (\ref{eq12}), the likelihood function for a sample of size n from this distribution is given by
\begin{equation}\label{eq24}
L(\theta_1, \theta_3) = \theta_1^n e^{-\theta_1\sum^n_{i=1}x_i}\theta_3^n(\prod^n_{i=1}(1+x_i)) e^{-\theta_3(\sum^n_{i=1}y_i+\sum^n_{i=1}x_iy_i)}
\end{equation}
This time we will consider a joint prior that is of the bivariate pseudo-gamma form. For it we assume that $\theta_3$ has a $\Gamma(\tau_1, \psi_1)$ density, i.e.,
\begin{equation*}
f(\theta_3) \propto \theta_3^{\tau_1-1}e^{-\theta_3\psi_1}I(\theta_3>0), ~~~~~ 0<\tau_1, \psi_1<\infty.
\end{equation*}
and then for each value of $\theta_3$, the conditional density $\theta_1$ given $\theta_3$ is assumed to be of the gamma form with an intensity parameter that is a linear function of $\theta_3$. Thus
\begin{equation*}
f(\theta_1|\theta_3) \propto (\psi_2+\psi_3\theta_3)^{\tau_2}\theta_1^{\tau_2-1}e^{-(\psi_2+\psi_3\theta_3)\theta_1}I(\theta_1>0).
\end{equation*}
where $0<\tau_2, \psi_3 <\infty$ and $0 \leq \psi_2 < \infty$.\\
The joint prior is thus of the form
\begin{equation}\label{eq25}
f(\theta_1,\theta_3) \propto (\psi_2+\psi_3\theta_3)^{\tau_2}\theta_1^{\tau_2-1}e^{-(\psi_2+\psi_3\theta_3)\theta_1}\theta_3^{\tau_1-1}e^{-\theta_3\psi_1}I(\theta_1>0,\theta_3>0).
\end{equation}
Consider the simpler prior in which we assume that $\psi_2=0$. This prior density will be of the form
\begin{equation}\label{eq26}
f_p(\theta_1, \theta_3) \propto \theta_1^{\tau_2-1}e^{-\theta_1\theta_3\psi_3}\theta_3^{\tau_1+\tau_2-1}e^{-\psi_1\theta_3}I(\theta_1>0, \theta_3>0).
\end{equation}
The corresponding posterior density to a sample of size n from sub-model I will be 
\begin{equation}\label{eq27}
\begin{split}
f(\theta_1,\theta_3|\underline{X}=\underline{x},\underline{Y}=\underline{y}) & \propto (\theta_1^n e^{-\theta_1\sum^n_{i=1}x_i})(\theta_3^n e^{-\theta_3(\sum^n_{i=1}y_i+\sum^n_{i=1}x_iy_i}) \\
& * (\theta_1^{\tau_2-1}e^{-\theta_1\theta_3\psi_3}\theta_3^{\tau_1+\tau_2-1}e^{-\psi_1\theta_3}) \\
& \propto \theta_1^{\tau_2+n-1}e^{-\theta_1\sum^n_{i=1}x_i}\theta_3^{\tau_1+\tau_2-1}e^{-\theta_3(\psi_1+\sum^n_{i=1}y_i+\sum^n_{i=1}x_iy_i)} \\
& * e^{-\theta_1\theta_3\psi_3} \\
& \propto (\theta_3^{\tau_1+n-1}e^{-\theta_3(\psi_1+\sum^n_{i=1}y_i+\sum^n_{i=1}x_iy_i)}) \\
& * (\theta_1^{\tau_2+n-1}e^{-\theta_1(\sum^n_{i=1}x_i+\psi_3\theta_3)})
\end{split}
\end{equation}
The marginal posterior distributions of $\theta_1$ and $\theta_3$ are
\begin{equation}\label{eq28}
f^p_{\theta_1}(\theta_1) \propto [\theta_1^{\tau_2+n-1}e^{-\theta_1\sum^n_{i=1}x_i}].\frac{\Gamma(\tau_1+n)}{(\psi_1+\sum^n_{i=1}y_i+\sum^n_{i=1}x_iy_i+\theta_1\psi_3)^{(\tau_1+n)}},
\end{equation}
and
\begin{equation}\label{eq29}
f^p_{\theta_1}(\theta_1) \propto [\theta_3^{\tau_1+n-1}e^{-\theta_3(\psi_1+\sum^n_{i=1}y_i+\sum^n_{i=1}x_iy_i)}].\frac{\Gamma(\tau_2+n)}{(\sum^n_{i=1}x_i+\theta_3\psi_3)^{(\tau_2+n)}}.
\end{equation}
For the plots of such prior and posterior densities, see Figures 4, 5, 6 (for priors) and Figures 7, 8, 9 (for posteriors).
\begin{figure}[htp]\label{fig4}
	\centering
	\includegraphics[width=6.5cm]{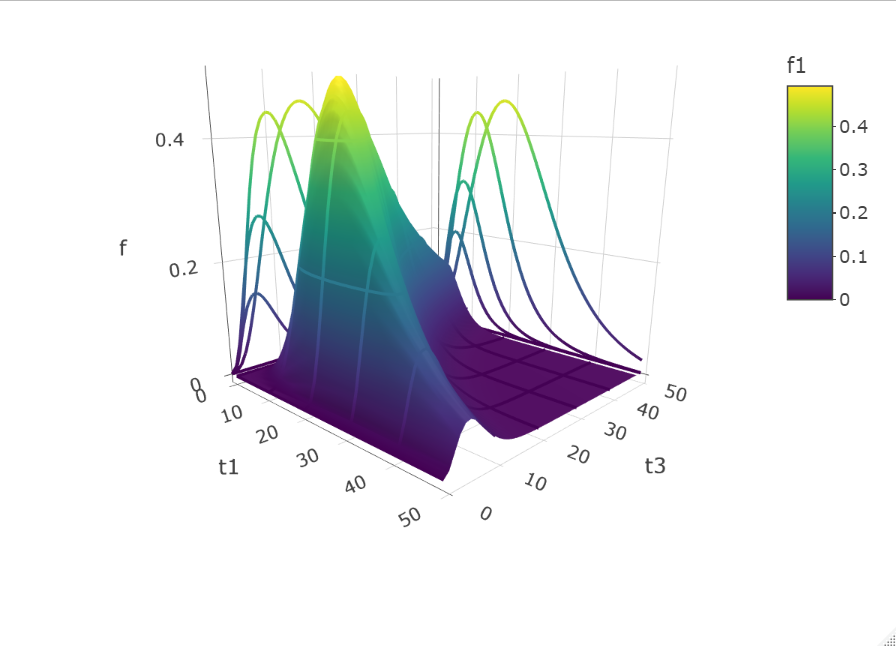}
	\caption{Density plot of pseudo-gamma prior with hyper-parameter values are $\tau_1=2, \tau_2=4, \psi_1=2, \psi_2=1(small), \psi_3=3$ and parameter values $\theta_1=2, \theta_3=5$ with sample of size 30 of the sub-model-I.}
	\label{fig:Density Plots}
\end{figure}

\begin{figure}[htp]\label{fig5}
	\centering
	\includegraphics[width=6.5cm]{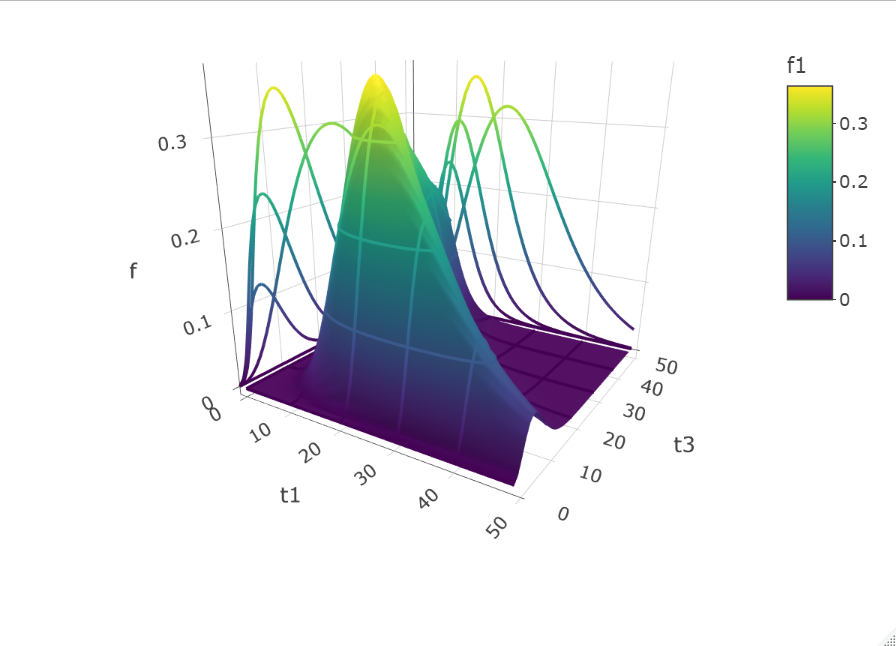}
	\caption{Density plot of pseudo-gamma prior with hyper-parameter values are $\tau_1=2, \tau_2=4, \psi_1=2, \psi_2=0(simple), \psi_3=3$ and parameter values $\theta_1=2, \theta_3=5$ with sample of size 30 of the sub-model-I.}
	\label{fig:Density Plots}
\end{figure}

\begin{figure}[htp]\label{fig6}
	\centering
	\includegraphics[width=6.5cm]{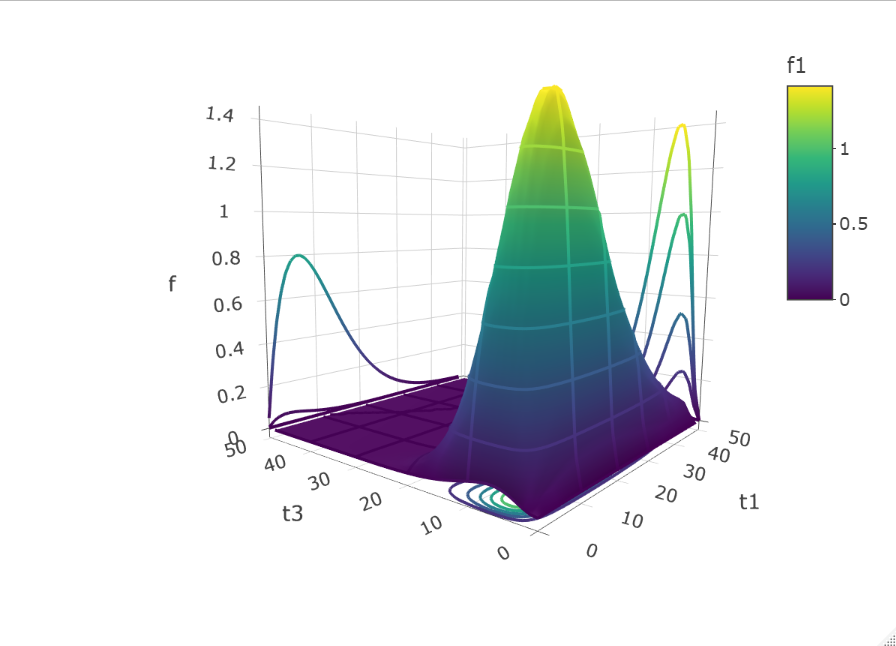}
	\caption{Density plot of pseudo-gamma prior with hyper-parameter values are $\tau_1=2, \tau_2=4, \psi_1=2, \psi_2=7(large), \psi_3=3$ and parameter values $\theta_1=2, \theta_3=5$ with sample of size 30 of the sub-model-I.}
	\label{fig:Density Plots}
\end{figure}

\begin{figure}[htp]\label{fig7}
	\centering
	\includegraphics[width=6.5cm]{figures/PDSMALL.png}
	\caption{Density plot of posterior (pseudo-gamma prior) with hyper-parameter values are $\tau_1=2, \tau_2=4, \psi_1=2, \psi_2=1(small), \psi_3=3$ and parameter values $\theta_1=2, \theta_3=5$ with sample of size 30 of the sub-model-I.}
	\label{fig:Density Plots}
\end{figure}

\begin{figure}[htp]\label{fig8}
	\centering
	\includegraphics[width=6.5cm]{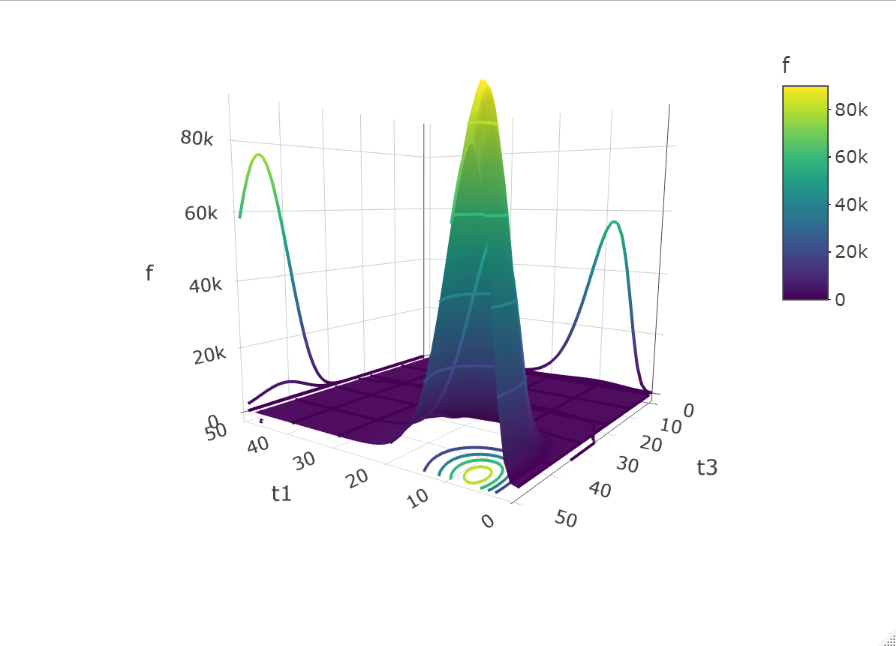}
	\caption{Density plot of posterior (pseudo-gamma prior) with hyper-parameter values are $\tau_1=2, \tau_2=4, \psi_1=2, \psi_2=0(simple), \psi_3=3$ and parameter values $\theta_1=2, \theta_3=5$ with sample of size 30 of the sub-model-I.}
	\label{fig:Density Plots}
\end{figure}

\begin{figure}[htp]\label{fig9}
	\centering
	\includegraphics[width=6.5cm]{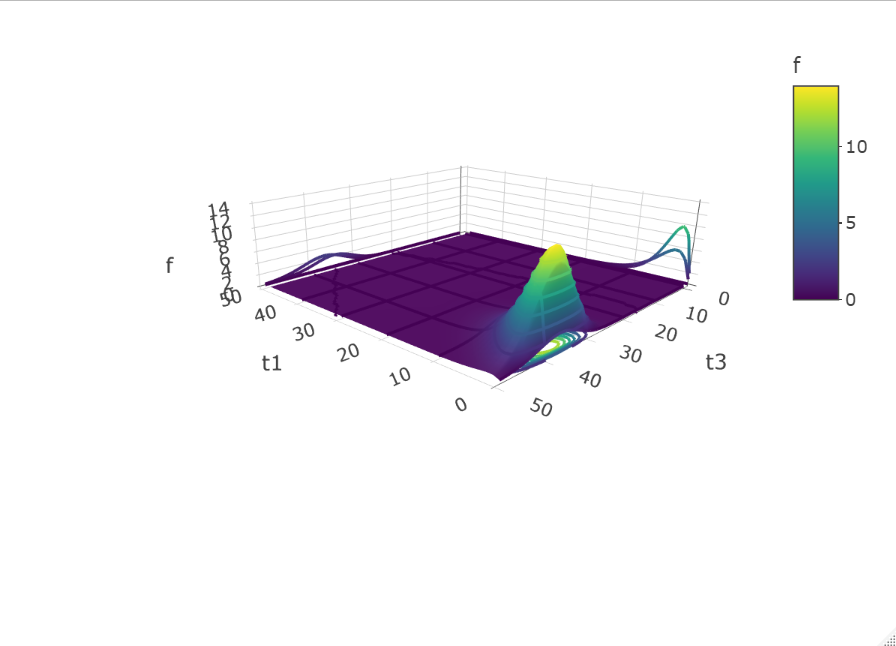}
	\caption{Density plot of posterior (pseudo-gamma prior) with hyper-parameter values are $\tau_1=2, \tau_2=4, \psi_1=2, \psi_2=7(large), \psi_3=3$ and parameter values $\theta_1=2, \theta_3=5$ with sample of size 30 of the sub-model-I.}
	\label{fig:Density Plots}
\end{figure}
\newpage
\subsection{Sub-Model-II ($\theta_2=0$)}
Consider the sub-model obtained by setting $\theta_2=0$. The model is thus of the form
\begin{equation*}
f_{X,Y}(x,y) = \theta_1e^{-\theta_1x}\theta_3xe^{-\theta_3xy}I(x>0,y>0).
\end{equation*}
The likelihood function for a sample of size n from this distribution is given by
\begin{equation*}
L(\theta_1, \theta_3) = \theta_1^n e^{-\theta_1\sum^n_{i=1}x_i}\theta_3^n(\prod^n_{i=1}x_i) e^{-\theta_3\sum^n_{i=1}x_iy_i}
\end{equation*}
the kernel of the likelihood function is given by
\begin{equation}\label{eq30}
L(\theta_1, \theta_3) \propto \theta_1^n e^{-\theta_1\sum^n_{i=1}x_i}\theta_3^ne^{-\theta_3\sum^n_{i=1}x_iy_i}
\end{equation}
Consider the simpler joint pseudo-gamma prior in which we assume that $\psi_2=0$. This joint prior density will be
\begin{equation}\label{eq31}
f_p(\theta_1, \theta_3) \propto \theta_1^{\tau_2-1}e^{-\theta_1\theta_3\psi_3}\theta_3^{\tau_1+\tau_2-1}e^{-\psi_1\theta_3}I(\theta_1>0, \theta_3>0).
\end{equation}
The posterior density will be
\begin{equation}\label{eq32}
\begin{split}
f(\theta_1,\theta_3|\underline{X}=\underline{x},\underline{Y}=\underline{y}) & \propto (\theta_1^n e^{-\theta_1\sum^n_{i=1}x_i})(\theta_3^n e^{-\theta_3(\sum^n_{i=1}x_iy_i}) \\
& * (\theta_1^{\tau_2-1}e^{-\theta_1\theta_3\psi_3}\theta_3^{\tau_1+\tau_2-1}e^{-\psi_1\theta_3}) \\
& \propto \theta_1^{\tau_2+n-1}e^{-\theta_1\sum^n_{i=1}x_i}\theta_3^{\tau_1+\tau_2-1} \\ &*e^{-\theta_3(\psi_1+\sum^n_{i=1}x_iy_i)}e^{-\theta_1\theta_3\psi_3} \\
& \propto (\theta_3^{\tau_1+n-1}e^{-\theta_3(\psi_1+\sum^n_{i=1}x_iy_i)}) \\
& * (\theta_1^{\tau_2+n-1}e^{-\theta_1(\sum^n_{i=1}x_i+\psi_3\theta_3)})
\end{split}
\end{equation}
The marginal posterior distributions of $\theta_1$ and $\theta_3$ are
\begin{equation}\label{eq33}
f^p_{\theta_1}(\theta_1) \propto [\theta_1^{\tau_2+n-1}e^{-\theta_1\sum^n_{i=1}x_i}].\frac{\Gamma(\tau_1+n)}{(\psi_1+\sum^n_{i=1}x_iy_i+\theta_1\psi_3)^{(\tau_1+n)}},
\end{equation}
and
\begin{equation}\label{eq34}
f^p_{\theta_1}(\theta_1) \propto [\theta_3^{\tau_1+n-1}e^{-\theta_3(\psi_1+\sum^n_{i=1}x_iy_i)}].\frac{\Gamma(\tau_2+n)}{(\sum^n_{i=1}x_i+\theta_3\psi_3)^{(\tau_2+n)}}.
\end{equation}
Note that the mean and variance of the marginals need to be dealt with numerically. 

\begin{figure}[htp]
	\centering
	\includegraphics[width=5.5cm]{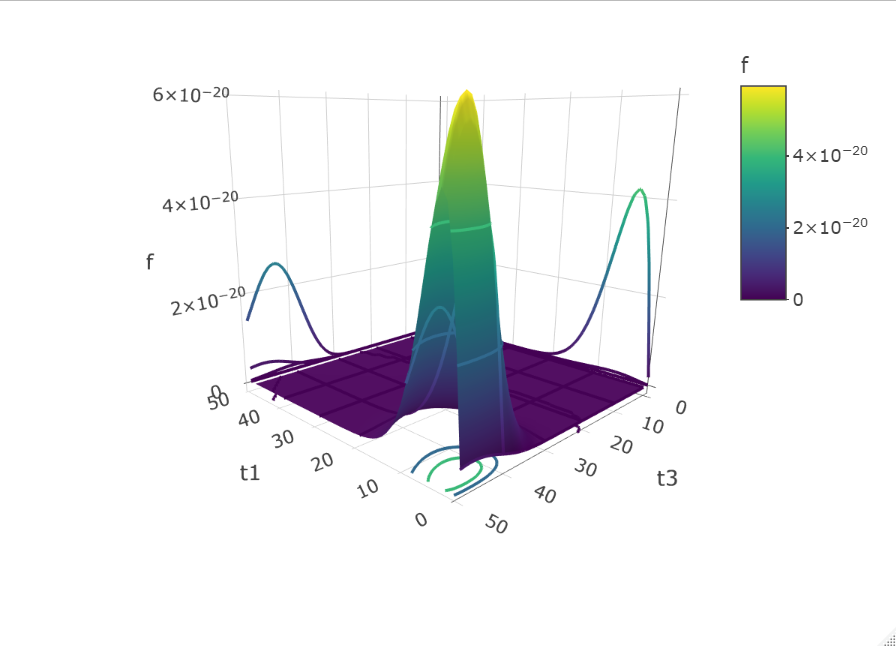}
	\caption{Density plot of posterior (pseudo-gamma prior) with hyper-parameter values are $\tau_1=2, \tau_2=4, \psi_1=2, \psi_2=1(small)$, $\psi_3=3$ and parameter values $\theta_1=2, \theta_3=5$ with sample of size 30 of the SM-II. }
	\label{fig:Density Plots}
\end{figure}

\begin{figure}[htp]
	\centering
	\includegraphics[width=5.5cm]{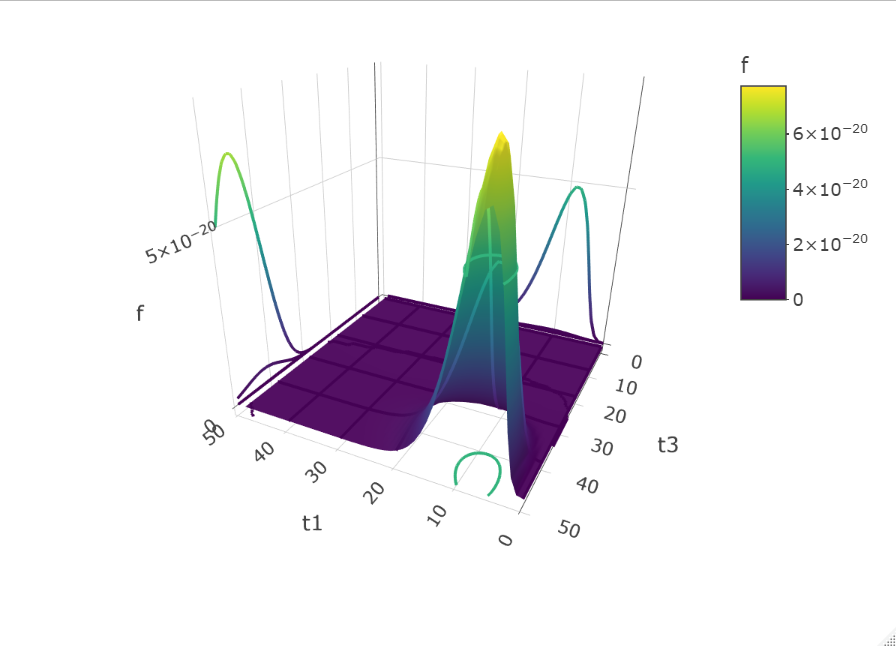}
	\caption{Density plot of posterior (pseudo-gamma prior) with hyper-parameter values are $\tau_1=2, \tau_2=4, \psi_1=2$,$ \psi_2=0(simple), \psi_3=3$ and parameter values $\theta_1=2, \theta_3=5$ with sample of size 30 of the SM-II.}
	\label{fig:Density Plots}
\end{figure}

\begin{figure}[htp]
	\centering
	\includegraphics[width=6.5cm]{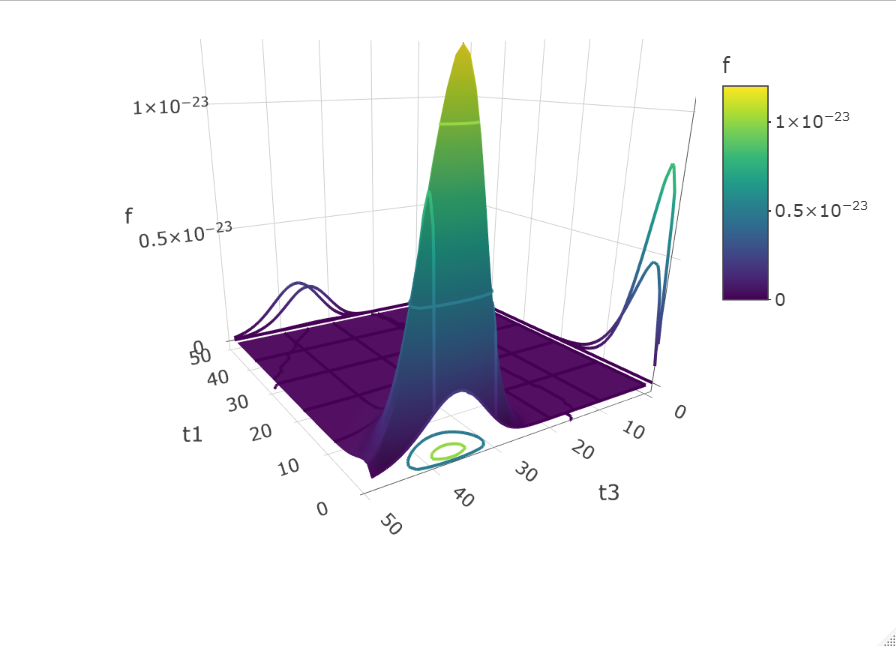}
	\caption{Density plot of posterior (pseudo-gamma prior) with hyper-parameter values are $\tau_1=2, \tau_2=4, \psi_1=2, \psi_2=7(large), \psi_3=3$ and parameter values $\theta_1=2, \theta_3=5$ with sample of size 30 of the SM-II.}
	\label{fig:Density Plots}
\end{figure}
\newpage

\section{Simulation study}
			Due to the general marginal and conditional composition of exponential distributions, simulation can be performed in two steps: first, simulate x from exponential($\theta_1$) and then y from exponential($\theta_2+\theta_3x$). However, even with independent gamma priors, inference for the posterior distribution is difficult for the full model. We must rely on a numerical algorithm to compute marginal distributions and their moments. In this paper, we will use the Hit-And-Run Metropolis (HARM) algorithm to simulate posterior distributions from observations for all priors (improper, independent, and pseudo) and for full and its sub-models. For the (HARM) algorithm implementation and comparison with Gibbs and Metropolis sampling, see  \cite{MHC1993}. We refer to \cite{BH2021} for current Bayesian simulation algorithms using R software.\\
			We have simulated 10,000 data sets using the HARM algorithm with thinning at
			10th sample from the posterior distributions of $\theta$’s with varying sample sizes of n =
			20, 30, 50, 100, 200, 500: Mean and posteriori 95\% confidence intervals are mentioned in Table 1 and 2 are computed from 10000 iterations of the preceding procedure.
			
			\begin{figure}[htp]
				\centering
				\includegraphics[width=7.5cm]{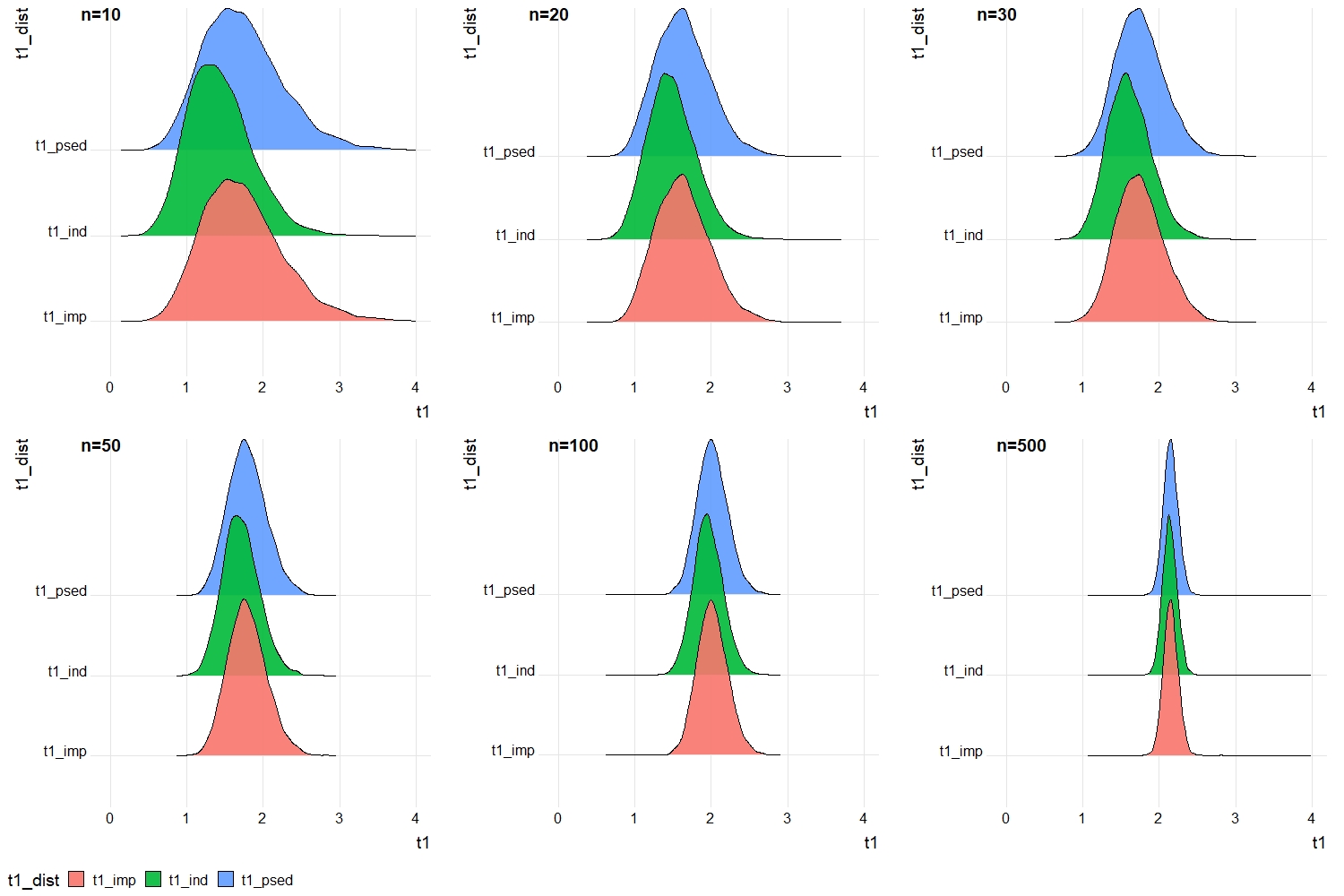}
				\caption{Posterior density plot of $\theta_1$ (independent gamma, improper and pseudo-gamma priors) with hyper-parameters $\alpha_1=2, \alpha_2=3, \alpha_3=4, \beta_1=2, \beta_2=1, \beta_3=5$ and parameter values $\theta_1=2, \theta_2=1, \theta_3=3$ with sample of size(n=10, 20, 30, 50, 100, 500).}
				\label{fig:Density Plots}
			\end{figure}
			
			\begin{figure}[htp]
				\centering
				\includegraphics[width=7.5cm]{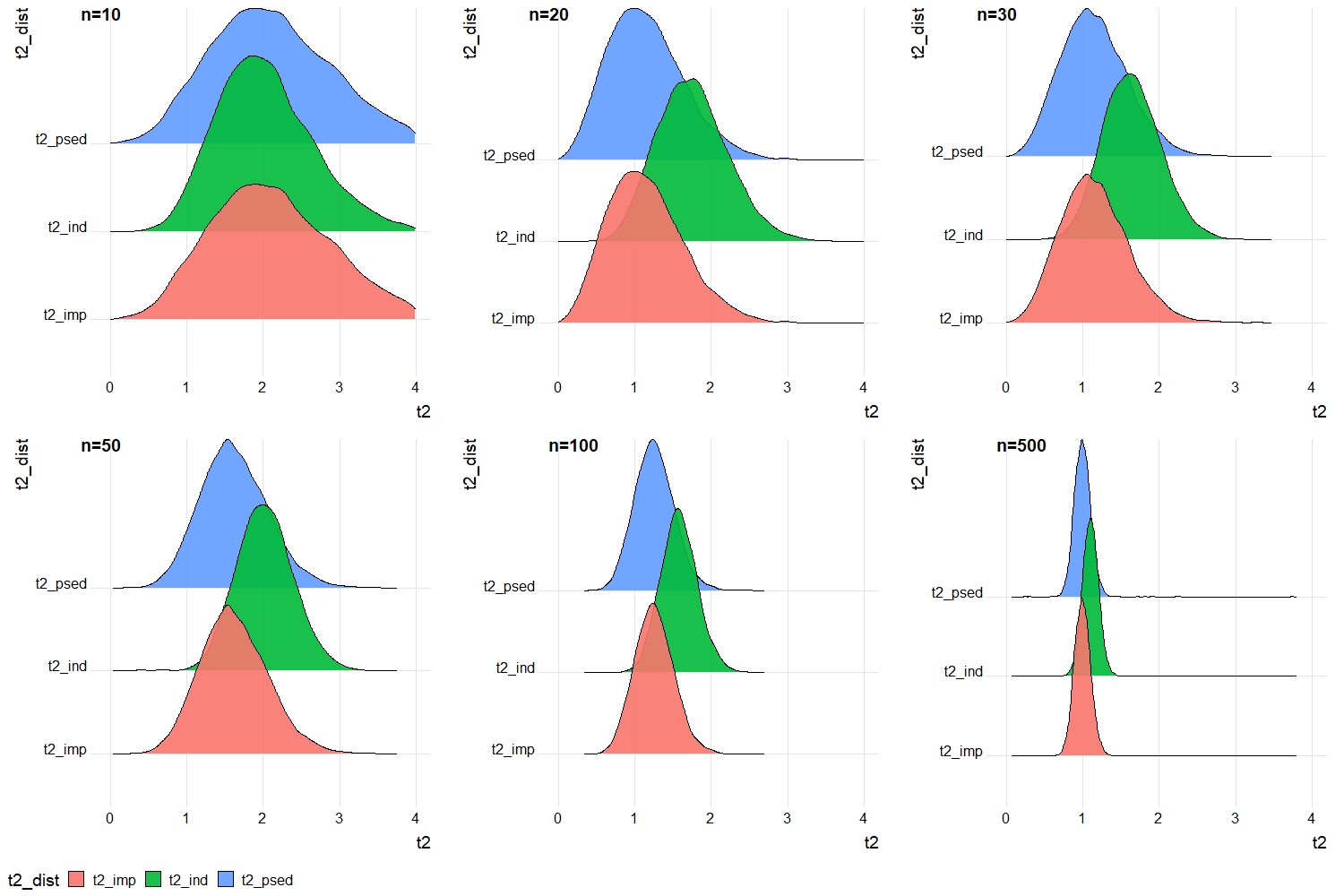}
				\caption{Posterior density plot of $\theta_2$ (independent gamma, improper and pseudo-gamma priors) with hyper-parameters $\alpha_1=2, \alpha_2=3, \alpha_3=4, \beta_1=2, \beta_2=1, \beta_3=5$ and parameter values $\theta_1=2, \theta_2=1, \theta_3=3$ with sample of size(n=10, 20, 30, 50, 100, 500).}
				\label{fig:Density Plots}
			\end{figure}
			\begin{figure}[htp]
				\centering
				\includegraphics[width=7.5cm]{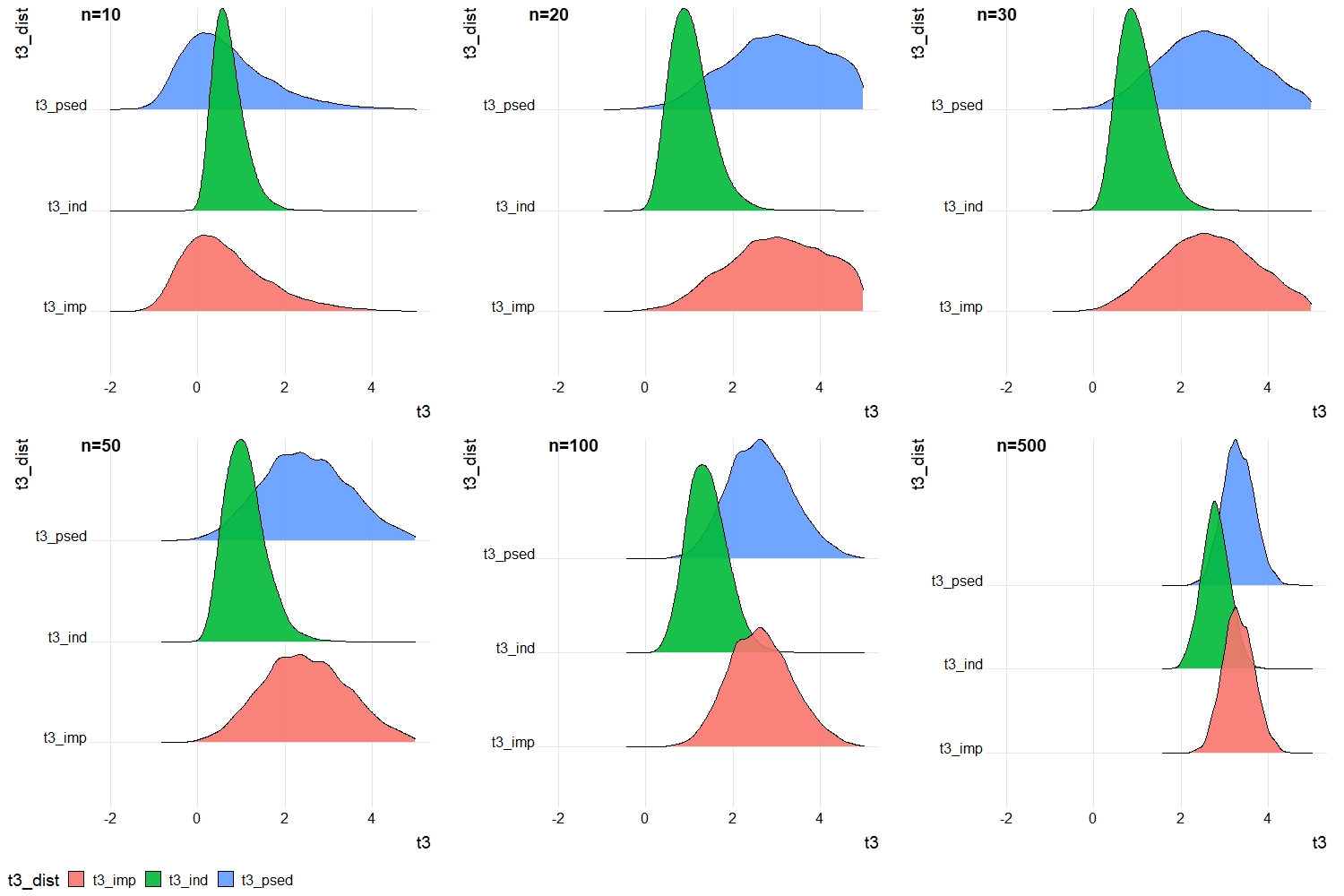}
				\caption{Posterior density plot of $\theta_3$ (independent gamma, improper and pseudo-gamma priors) with hyper-parameters $\alpha_1=2, \alpha_2=3, \alpha_3=4, \beta_1=2, \beta_2=1, \beta_3=5$ and parameter values $\theta_1=2, \theta_2=1, \theta_3=3$ with sample of size(n=10, 20, 30, 50, 100, 500).}
				\label{fig:Density Plots}
			\end{figure}
			\begin{figure}[htp]
				\centering
				\includegraphics[width=7.5cm]{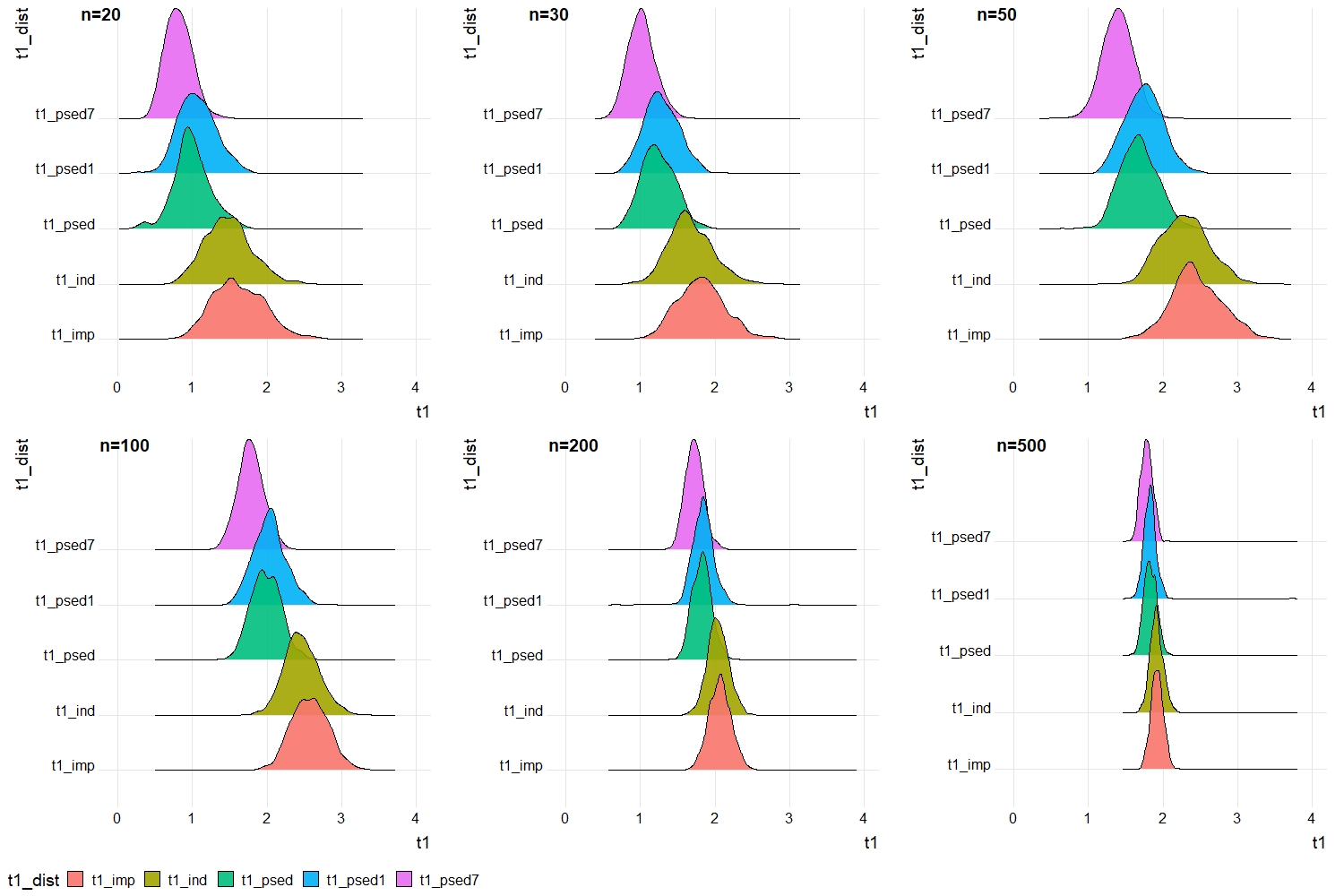}
				\caption{Posterior density plot of $\theta_1$  with hyper-parameters $\alpha_1=2, \alpha_3=4, \beta_1=2, \beta_3=5$ and parameter values $\theta_1=2, \theta_3=3$ of sub-model-I.}
				\label{fig:Density Plots}
			\end{figure}
			
			\begin{figure}[htp]
				\centering
				\includegraphics[width=7.5cm]{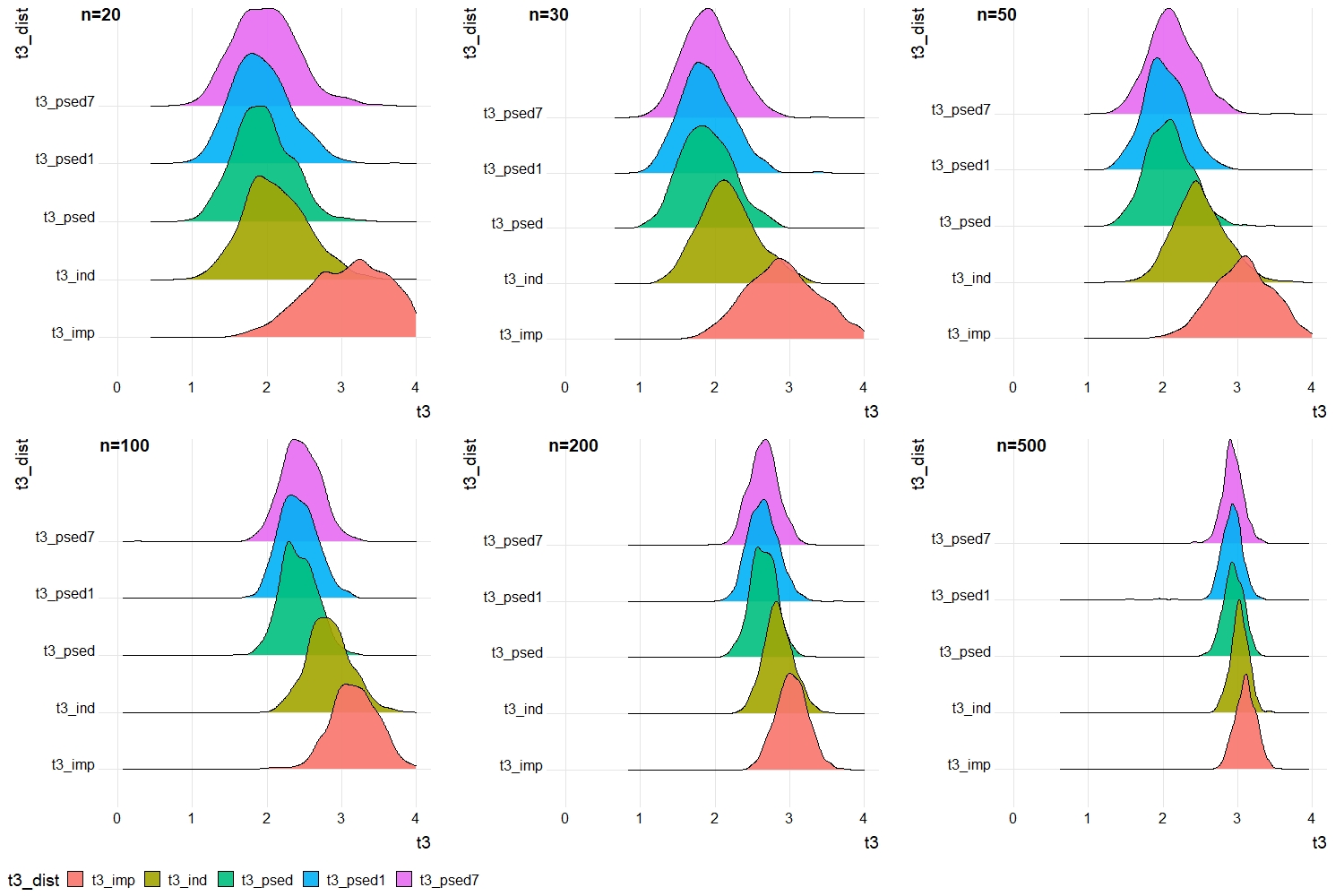}
				\caption{Posterior density plot of $\theta_3$  with hyper-parameters $\alpha_1=2, \alpha_3=4, \beta_1=2, \beta_3=5$ and parameter values $\theta_1=2, \theta_3=3$ of sub-model-I.}
				\label{fig:Density Plots}
			\end{figure}
			
			\begin{figure}[htp]
				\centering
				\includegraphics[width=7.5cm]{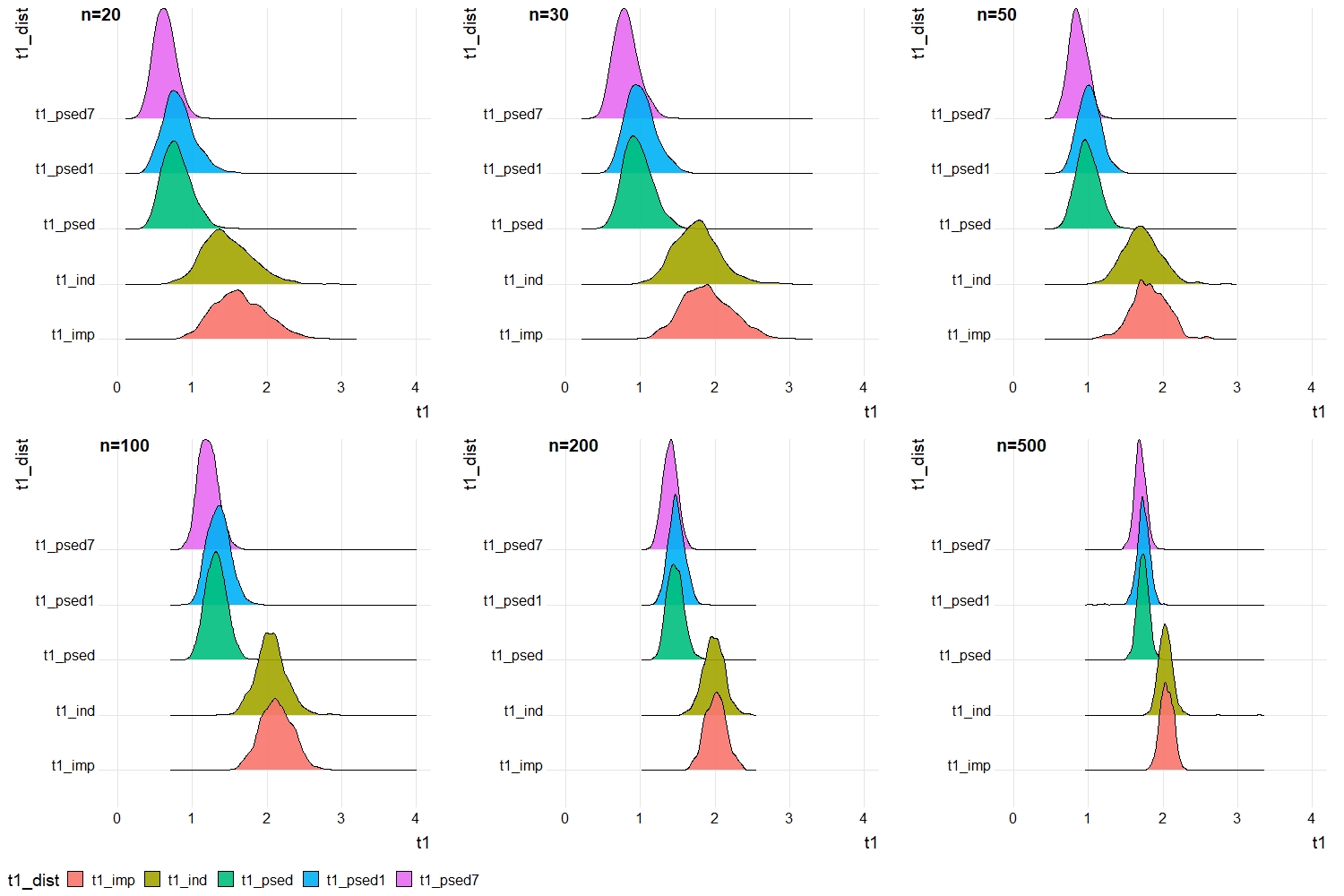}
				\caption{Posterior density plot of $\theta_1$  with hyper-parameters $\alpha_1=2, \alpha_3=4, \beta_1=2, \beta_3=5$ and parameter values $\theta_1=2, \theta_3=3$ of sub-model-II.}
				\label{fig:Density Plots}
			\end{figure}
			
			\begin{figure}[htp]
				\centering
				\includegraphics[width=7.5cm]{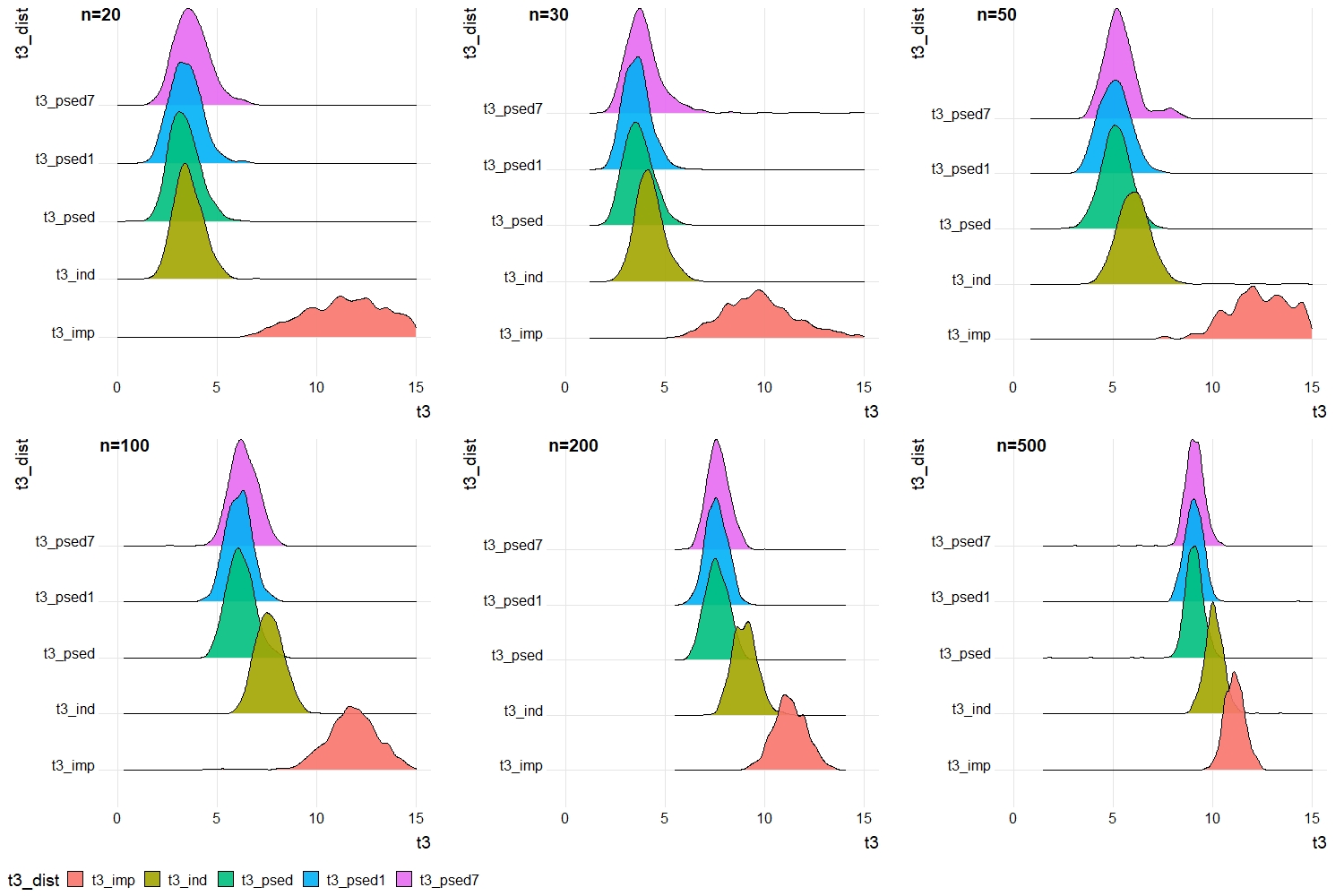}
				\caption{Posterior density plot of $\theta_1$  with hyper-parameters $\alpha_1=2, \alpha_3=4, \beta_1=2, \beta_3=5$ and parameter values $\theta_1=2, \theta_3=3$ of sub-model-II.}
				\label{fig:Density Plots}
			\end{figure}

			\begin{landscape}
				\begin{table}
					\caption{{ Simulation (Sub-model I)}} 
					\label{Table_sub_I}
					\normalsize
					\begin{tablenotes}
						\item {\bf Sample size (n), parameters (P), posterior mean by independent gamma prior (IGP(1)), posterior mean by improper prior (ImP(2)), posterior mean by pseudo-gamma prior for $\psi_2=1$ (PGP1(3)), posterior mean by pseudo-gamma prior for $\psi_2=0$ (PGP2(4)), posterior mean by pseudo-gamma prior for $\psi_2=7$ (PGP3(5)),  95\% confidence interval of IGP(1) (CI(1)), 95\% confidence interval of ImP(2) (CI(2)), 95\% confidence interval of PGP1(3) (CI(3)), 95\% confidence interval of PGP2(4) (CI(4)), 95\% confidence interval of PGP3(5) (CI(5))}
					\end{tablenotes}
					\centering 
					\begin{tabular}{lcccccccccccccccccr} 
						\toprule[\heavyrulewidth]\toprule[\heavyrulewidth]
						\textbf{$n$ }  & \textbf{P} & \textbf{SIP$_1$} & \textbf{SIP$_2$} & \textbf{SIP$_3$} & \textbf{SIP$_4$} &\textbf{SIP$_5$} & \textbf{CISIP$_1$} &  \textbf{CISIP$_2$}   &  \textbf{CISIP$_3$}   &   \textbf{CISIP$_4$} & \textbf{CISIP$_5$}   \\ 
						\midrule
						
						\multirow{2}{*}{$20$} & $\theta_1$ & 1.483 & 1.624 & 0.998 &  1.071 &  0.815  & (0.924 2.141) & (1.015 2.371) & (0.504 1.550) & (0.608 1.608) & (0.516 1.172) \\
						& $\theta_3$ & 2.304 & 3.306 & 2.239 &  2.075 &  1.984 &  (1.350 3.291) & (2.028 4.965) & (1.257 9.249) & (1.258 2.935) & (1.259 2.936)\\
						\bottomrule[\heavyrulewidth]    
						
						\multirow{2}{*}{$30$} & $\theta_1$ & 1.729 & 1.953 & 1.242 &  1.287 &  1.016 &  (1.152 2.365) & (1.257 2.648) & (0.836 1.707) & (0.853 1.769) & (0.702 1.401) \\
						& $\theta_3$ & 2.184 & 3.159 & 1.895 &  1.884 &  1.929 &  (1.486 3.041) & (2.011 4.498) & (1.282 2.643) & (1.283 2.616) & (1.319 2.625)\\
						\bottomrule[\heavyrulewidth]    
						
						\multirow{2}{*}{$50$} & $\theta_1$ & 2.294 & 2.450 & 1.677 &  1.746 &  1.404 &  (1.723 2.926) & (1.813 3.127) & (1.240 2.179) & (1.273 2.261) & (0.984 1.808) \\
						& $\theta_3$ & 2.488 & 3.091 & 2.121 &  2.030 &  2.309 &  (1.845 3.193) & (2.248 4.013) & (1.517 2.765) & (1.482 2.621) & (1.547 2.868)\\
						\bottomrule[\heavyrulewidth]    
						
						\multirow{2}{*}{$100$} & $\theta_1$ & 2.465 & 2.918 & 1.986  & 2.032 &  1.782 &  (2.033 2.981) & (2.129 3.265) & (1.625 2.399) & (1.648 2.479) & (1.449 2.128) \\
						& $\theta_3$ & 2.830 & 3.175 & 2.422 &  2.417 &  2.450 &  (2.278 3.435) & (2.565 3.811) & (1.970 2.892) & (1.971 2.942) & (1.961 2.973)\\
						
						\bottomrule[\heavyrulewidth]    
						
						\multirow{2}{*}{$200$} & $\theta_1$ & 2.039 & 2.066 & 1.823 &  1.845 &  1.801 &  (1.765 2.324) & (1.776 2.362) & (1.582 2.088) & (1.588 2.132) & (1.503 2.020) \\
						& $\theta_3$ & 2.850 & 3.028 & 2.650 &  2.813 &  2.663 &  (2.474 3.258) & (2.625 3.425) & (2.281 3.010) & (2.303 3.180) & (2.279 3.051)\\
						
						\bottomrule[\heavyrulewidth]
						
						\multirow{2}{*}{$500$} & $\theta_1$ & 1.912 & 1.944 & 1.828 &  2.245 &  1.781 &  (1.747 2.096) & (1.761 2.085) & (1.673 1.996) & (1.669 2.042) & (1.639 1.934) \\
						& $\theta_3$ & 3.026 & 3.104 & 2.944 &  2.906  & 2.934 &  (2.776 3.262) & (2.829 3.384) & (2.676 3.206) & (2.614 3.190) & (2.669 3.204)\\
						\bottomrule[\heavyrulewidth]    
						\bigskip
						
					\end{tabular}
				\end{table}
			\end{landscape}
			
			\begin{landscape}
				\begin{table}
					\caption{{ Simulation (Sub-model II)}} 
					\label{Table_sub_II}
					\normalsize
					\begin{tablenotes}
						\item {\bf Sample size (n), parameters (P), posterior mean by independent gamma prior (IGP(1)), posterior mean by improper prior (ImP(2)), posterior mean by pseudo-gamma prior for $\psi_2=1$ (PGP1(3)), posterior mean by pseudo-gamma prior for $\psi_2=0$ (PGP2(4)), posterior mean by pseudo-gamma prior for $\psi_2=7$ (PGP3(5)),  95\% confidence interval of IGP(1) (CI(1)), 95\% confidence interval of ImP(2) (CI(2)), 95\% confidence interval of PGP1(3) (CI(3)), 95\% confidence interval of PGP2(4) (CI(4)), 95\% confidence interval of PGP3(5) (CI(5))}
					\end{tablenotes}
					\centering 
					\begin{tabular}{lcccccccccccccccccr} 
						\toprule[\heavyrulewidth]\toprule[\heavyrulewidth]
						\textbf{$n$ }  & \textbf{P} & \textbf{SIP$_1$} & \textbf{SIP$_2$} & \textbf{SIP$_3$} & \textbf{SIP$_4$} &\textbf{SIP$_5$} & \textbf{CISIP$_1$} &  \textbf{CISIP$_2$}   &  \textbf{CISIP$_3$}   &   \textbf{CISIP$_4$} & \textbf{CISIP$_5$}   \\ 
						\midrule
						
						\multirow{2}{*}{$20$} & $\theta_1$ & 1.473 & 1.643 & 0.792 & 0.822 &  0.629 & (0.900 2.177) & (1.040 2.394) & (0.481 1.192) & (0.469 1.321) & (0.383 0.913) \\
						& $\theta_3$ & 3.744 & 11.814 & 3.344 & 3.417 &  3.739 & (2.297 5.415) & (7.516 16.266) & (2.065 4.996) & (2.045 5.184) & (2.227 5.714)\\
						\bottomrule[\heavyrulewidth]    
						
						\multirow{2}{*}{$30$} & $\theta_1$ & 1.757 & 1.909 & 0.969 & 1.001 &  1.103 & (1.222 2.399) & (1.268 2.617) & (0.633 1.399) & (0.667 1.444) & (0.540  1.400) \\
						& $\theta_3$ & 4.207 & 9.799 & 3.679 & 3.565 &  4.066 & (2.954 5.829) & (6.403 14.006) & (2.452 5.282) & (2.430 4.943) & (2.538 6.711)\\
						\bottomrule[\heavyrulewidth]    
						
						\multirow{2}{*}{$50$} & $\theta_1$ & 1.824 & 1.821 & 0.988 & 1.011 &  0.860 & (1.295 2.251) & (1.337 2.247) & (0.718 1.295) & (0.738 1.329) & (0.644 1.070) \\
						& $\theta_3$ & 6.041 & 12.405 & 5.126 &  5.070 &  5.388 & (4.452 7.781) & (9.154 14.984) & (3.653 6.581) & (3.689 6.629) & (4.063 7.842)\\
						\bottomrule[\heavyrulewidth]    
						
						\multirow{2}{*}{$100$} & $\theta_1$ & 2.057 & 2.285 & 1.316 & 1.351 &  1.210 & (1.680 2.481) & (1.711 2.646) & (1.052 1.604) & (1.087 1.649) & (0.982 1.478) \\
						& $\theta_3$ & 7.628 & 11.732 & 6.154 & 6.064 &  6.306 & (6.319 9.046) & (9.091 14.179) & (4.912 7.569) & (4.872 7.438) & (4.984 7.669)\\
						
						\bottomrule[\heavyrulewidth]    
						
						\multirow{2}{*}{$200$} & $\theta_1$ & 1.983 & 2.005 & 1.466 & 1.482 &  1.398 &  (1.709 2.265) & (1.730  2.290) & (1.271 1.694) & (1.276 1.690) & (1.208 1.592) \\
						& $\theta_3$ & 8.995 & 11.257 & 7.583 &  7.565  & 7.661 & (7.880 10.261) & (9.636 12.890) & (6.516 8.689) & (6.486 8.579) & (6.631 8.792)\\
						
						\bottomrule[\heavyrulewidth]
						
						\multirow{2}{*}{$500$} & $\theta_1$ & 2.212 & 2.048 & 1.731 & 1.732 &  1.694 & (1.866 2.270) & (1.876 2.211) & (1.589 1.880) & (1.558 1.894) & (1.543 1.848) \\
						& $\theta_3$ & 10.185 & 11.098 & 9.047 & 9.253 &  9.083 & (9.129 11.412) & (10.163 12.109) & (8.268 9.915) & (8.177 10.141) & (8.301 10.008)\\
						\bottomrule[\heavyrulewidth]    
						\bigskip
						
					\end{tabular}
				\end{table}
			\end{landscape}

			\newpage

\section{Applications:}
One example from the social sciences where an inverse dependence relationship is expected is infant mortality and GDP. Both of these indicators are roughly exponentially distributed for countries and regions all over the world. The C.I.A. generates estimates of values based on massive databases of global affairs information. In 2022, 225 countries and regions will provide data on infant mortality as deaths per 1000 live births (Y) and per capita GDP in thousands of dollars (X).
\begin{figure}[htp]
	\centering
	\includegraphics[width=8cm]{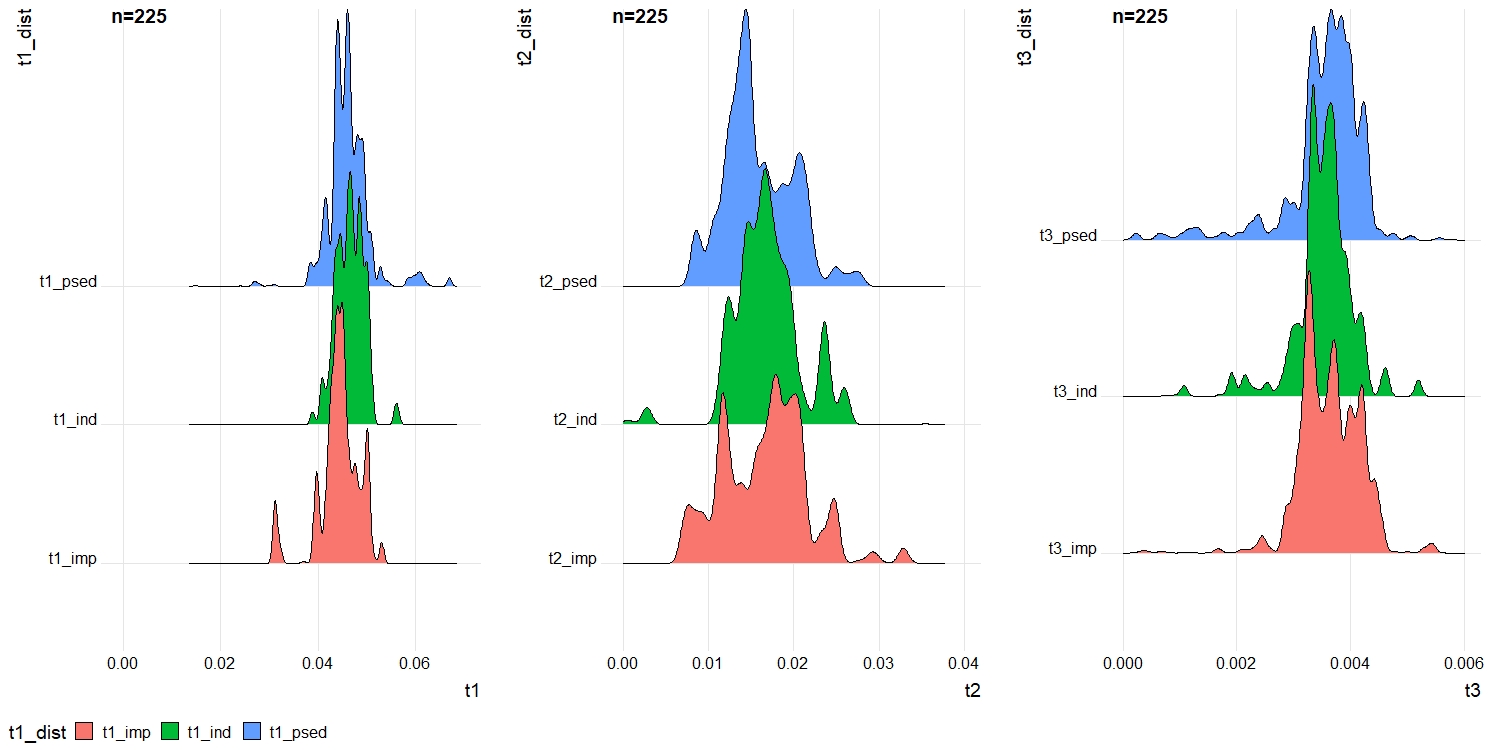}
	\caption{Posterior density plots for full-model}
	\label{fig:Density Plots}
\end{figure}

\begin{figure}[htp]
	\centering
	\includegraphics[width=7.5cm]{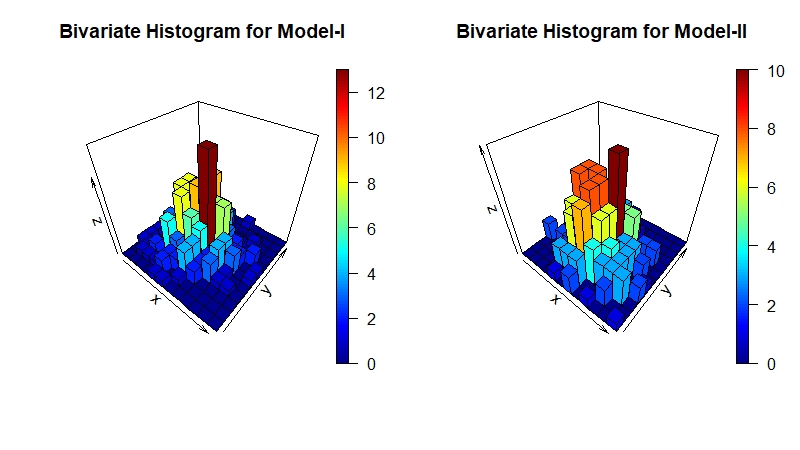}
	\caption{Histograms for the bivariate data of the per Capita GDP and Infant Mortality rate data set for sub-model-I \& II}
	\label{fig:Density Plots}
\end{figure}

		\begin{table}
	\caption{{per Capita GDP and Infant Mortality rate dataset (full-model)}} 
	\label{Table_sub_I}
	\normalsize
	\centering 
	\begin{tabular}{lccccccccr} 
		\toprule[\heavyrulewidth]\toprule[\heavyrulewidth]
		\textbf{n}  & \textbf{P} & \textbf{IGP(1)} & \textbf{ImP(2)} & \textbf{PGP1(3)} &  \textbf{CI(1)} &  \textbf{CI(2)}   &  \textbf{CI(3)}    \\ 
		\midrule
		
		\multirow{3}{*}{$225$} & $\theta_1$ & 0.209 & 0.113 & 0.112 & (0.041 0.056) & (0.031 0.170) & (0.040 0.321) \\
		& $\theta_2$ & 0.026 & 0.032 & 2.234 & (0.011 0.056) & (0.007 0.033) & (0.009 35.763) \\
		& $\theta_3$ & 0.004 & 0.098 & 0.022 & (0.002 0.005) & (0.003 0.005) & (0.001 0.016) \\
		\bottomrule[\heavyrulewidth]    
		
		
	\end{tabular}
	
\end{table}
\begin{table}
	\caption{{per Capita GDP and Infant Mortality rate dataset (sub-model-I)}} 
	\label{Table_sub_I}
	\normalsize
	\centering 
	\resizebox{\textwidth}{!}{
		\begin{tabular}{lccccccccr} 
			\toprule[\heavyrulewidth]\toprule[\heavyrulewidth]
			\textbf{n}  & \textbf{P} & \textbf{IGP(1)} & \textbf{ImP(2)} & \textbf{PGP1(3)} &  \textbf{CI(1)} &  \textbf{CI(2)}   &  \textbf{CI(3)}    \\ 
			\midrule
			
			\multirow{2}{*}{$225$} & $\theta_1$ & 0.04103 & 0.04739 & 0.05478 & (0.04906, 0.04113) & (0.04729, 0.04773) & (0.05410, 0.05768) \\
			
			& $\theta_3$ & 0.00468 & 0.00451 & 0.00420 & (0.00447, 0.00498) & (0.00441, 0.00490) & (0.00400, 0.00464) \\
			\bottomrule[\heavyrulewidth]    
			
			
		\end{tabular}
	}
	
\end{table}

\begin{table}
	\caption{{per Capita GDP and Infant Mortality rate dataset (sub-model-II)}} 
	\label{Table_sub_I}
	\normalsize
	\centering 
	\resizebox{\textwidth}{!}{
		\begin{tabular}{lccccccccr} 
			\toprule[\heavyrulewidth]\toprule[\heavyrulewidth]
			\textbf{n}  & \textbf{P} & \textbf{IGP(1)} & \textbf{ImP(2)} & \textbf{PGP1(3)} &  \textbf{CI(1)} &  \textbf{CI(2)}   &  \textbf{CI(3)}    \\ 
			\midrule
			
			\multirow{2}{*}{$225$} & $\theta_1$ & 0.04131 & 0.05280 & 0.04991 & (0.04125, 0.04142) & (0.04013, 0.06238) & (0.04768, 0.05091) \\
			
			& $\theta_3$ & 0.00506 & 0.00484 & 0.00508 & (0.00484, 0.00536) & (0.00395, 0.00588) & (0.00481, 0.00550) \\
			\bottomrule[\heavyrulewidth]    
			
			
		\end{tabular}
	}
	
\end{table}

\section{Conclusion}		
\cite{BM2019} have argued that the pseudo-exponential distribution should be the primary option when modeling bivariate data. The pseudo-exponential provides for both classical and Bayesian analysis as well as simple simulation due to its clear construction in \cite{bbb2022}. However, only a tiny number of bivariate priors in the literature can account for dependence between the parameters when there is outside information indicating it. We introduced the pseudo-gamma priors in Section 4 and suggested as a possible option for Bayesian inference with prior dependence. Finally recommend that Bayesian analysis with dependent prior  for the bivariate  pseudo-exponential with a pseudo-gamma priors.
\section{Acknowledgement(s)}
I thank Ministry of Tribal Affairs-National Fellowship Scheme For Higher Education of ST Students(NFST) for providing me Junior Research Fellowship (JRF) and Senior Research Fellowship (SRF) (award no: 201819-NFST-TEL-00347). Also thanks to Institution of Eminence(IoE) project, University of Hyderabad, Hyderabad (UoH-IoE-RC2-21-013).

		\end{document}